\documentclass[journal]{IEEEtran}

\usepackage{pifont}
\usepackage{amsfonts}
\usepackage[cmex10]{amsmath}
\usepackage{stmaryrd}
\usepackage{bbding}
\usepackage{txfonts}
\usepackage{amssymb}
\usepackage{graphicx}
\usepackage{subfigure}

\usepackage{multirow}
\usepackage{algorithm}
\usepackage{algorithmic}
\usepackage{bm}
\usepackage{paralist}
\usepackage{cite}
\usepackage{url}
\usepackage{color}
\usepackage{ulem}

\usepackage{cases}
\usepackage{balance}
\usepackage{array}
\usepackage{stfloats}
\usepackage{flushend}

\begin{document}
\title{Deep Learning-Based Detection for Marker Codes over Insertion and Deletion Channels}%
\author{Guochen Ma, Xiaopeng Jiao, Jianjun Mu, Hui Han, and Yaming Yang
\thanks{This work is supported by the National Natural Science Foundation of China under Grants 61971322, 61977051, and 62001362.}
\thanks{The authors are with the School of Computer Science and Technology, Xidian University, Xi'an 710071, China. (e-mail: guochenma@163.com; jiaozi1216@126.com; jjmu@xidian.edu.cn; huihan0424@163.com; yym@xidian.edu.cn).)}%

}

\maketitle

\begin{abstract}
Marker code is an effective coding scheme to protect data from insertions and deletions. It has potential applications in future storage systems, such as DNA storage and racetrack memory. When decoding marker codes, perfect channel state information (CSI), i.e., insertion and deletion probabilities, are required to detect insertion and deletion errors. Sometimes, the perfect CSI is not easy to obtain or the accurate channel model is unknown. Therefore, it is deserved to develop detecting algorithms for marker code without the knowledge of perfect CSI. In this paper, we propose two {CSI-agnostic} detecting algorithms for marker code based on deep learning. The first one is a model-driven deep learning method, which deep unfolds the original iterative detecting algorithm of marker code. In this method, CSI become weights in neural networks and these weights can be learned from training data. The second one is a data-driven method which is an end-to-end system based on the deep bidirectional gated recurrent unit network. Simulation results show that error performances of the proposed methods are significantly better than that of the original detection algorithm with CSI uncertainty. Furthermore, the proposed data-driven method exhibits better error performances than other methods for unknown channel models.

\end{abstract}


\IEEEpeerreviewmaketitle

\vspace{-10pt}

\section{Introduction} \label{introduction}

\IEEEPARstart{C}{hannels} with synchronization errors \cite{Mercier10}, including insertions and deletions, have appeared in various communication and storage systems, such as wireless communication \cite{Tinnirello16}\cite{Chen16}, racetrack memory \cite{Chee18}, and DNA storage \cite{Shomorony22}. Synchronization errors can cause catastrophic consequences for communication systems. Even a single insertion/deletion error makes quite a few bits different between the transmitter and the receiver. Traditional error correcting codes for substitution errors cannot be used in this scenario. Thus, constructing efficient codes against insertions and deletions has received a lot of research interest.

For channels with fixed number of insertion/deletion errors, zero-error number-theoretic codes have been studied extensively. The well-known Varshamov-Tenengolts (VT) code, invented several decades ago, is a single insertion/deletion error correcting code with asymptotically optimal redundancy \cite{Levenshtein65}. Helberg \textit{et al.} extended VT codes to multiple insertion/deletion errors via a modified Fibonacci sequence, but the coding rate is rather low \cite{Helberg02}. More recently, number-theoretic codes correcting two deletions with low redundancy have been successfully constructed \cite{Gabrys19}\cite{Sima20}\cite{Guruswami21}. However, the decoding algorithm becomes complicated when extending the code constructed in \cite{Sima20} to more than two insertion/deletion errors \cite{Sima21}.

When relaxing the adversarial noise assumption and zero-error requirement, probabilistic synchronization channels and code constructions that tolerate a small probability of errors are studied, which often lead to practical solutions for multiple insertion/deletion errors. Davey \textit{et al.} introduced a random insertions, deletions, and substitutions (IDS) channel model and designed a concatenated scheme using an inner watermark code and an outer low-density parity-check (LDPC) code over a nonbinary field \cite{Davey01}. The synchronization is recovered using a forward-backward (FB) algorithm via a pseudorandom watermark sequence shared by the transmitter and the receiver. Later, Ratzer proposed marker codes for IDS channels which use periodic markers inserted into the information sequence as inner codes and binary LDPC codes as outer codes \cite{Ratzer05}. The synchronization of marker codes is also recovered by the FB algorithm via markers. Compared with watermark codes, marker codes sometimes exhibit lower decoding complexity and better error rate performances. The methods of the above two codes for IDS channels have also been improved and extended to other scenarios in recent years \cite{Briffa10,Jiao11,Wang11,Jiao12,Yazdani22,Han13,Goto18,Shibata22,Achari23}.

In this paper, we investigate a decoding problem of marker codes for insertion and deletion channels that is not considered in the related literature before. When decoding marker codes or other similar codes for insertion and deletion channels, the channel state information (CSI), such as insertion and deletion probabilities, should be known perfectly at the receiver. Then the FB algorithm can be used to recover synchronization optimally for each transmitted symbol. However, the performance of FB algorithm for synchronization recovery is significantly degraded in the presence of CSI uncertainty, i.e., the insertion and deletion probabilities are not known accurately at the receiver. Moreover, when the accurate channel model is unknown, e.g., the insertion and deletion events are not random and independent, the FB algorithm may not be optimal. Motivated by the application of deep learning for physical layer communications \cite{Wang17,OShea17,Dorner18,Simeone18,Qin19,He19}, we propose deep learning-based detection methods for marker code over insertion and deletion channels to solve the performance degradation problem due to CSI uncertainty or inaccurate channel models. {Our study is also driven by the complexity of DNA storage channel model related to insertion/deletion errors. There are various kinds of errors in DNA storage systems, including insertions, deletions, substitutions, and sequence losses \cite{Heckel19}. The errors are not random, and sometimes burst, during DNA synthesis and DNA sequencing \cite{Heckel19,Donnell13,Yazdi15}. This may lead to inaccurate estimation of CSI or even inaccurate characterization of the DNA storage channel model. }

\subsection{Related Work on Deep Learning for Channel Detection {and Decoding} }

{ In recent years, deep learning-based detection and decoding methods for digital communication systems have received a lot of attention. In general, these deep learning-based methods can be classified into black-box (data-driven) approaches and model-based approaches \cite{Shlezinger23}. }

\subsubsection{\textbf{{Black-box approaches}}} {In these approaches, the channel detection/decoding problems are solved directly by classical deep learning models. They do not rely on the model-based detection/decoding algorithms developed in the digital communication field.}

For detecting Poisson channel used in optical and molecular communication systems, Farsad \textit{et al.} proposed a data-driven sliding bidirectional recurrent neural network (SBRNN) detector \cite{Farsad18}. It performs well without any knowledge of the underlying channel model. When the channel model is known, the SBRNN detector without the need of CSI performs much better than the well-known Viterbi detector with imperfect CSI. For detecting multiple-input multiple-output (MIMO) systems, Samuel \textit{et al.} in \cite{Samuel19} proposed two deep learning-based detectors, one of them is a black-box method which uses a standard fully connected multi-layer network. For magnetic recording channels with intersymbol interference (ISI), the authors in \cite{Zheng21} proposed a recurrent neural network (RNN)-based detection method, named partial-response neural network (PR-NN). The bi-directional gated recurrent units (bi-GRUs) are used for PR-NN to recover the ISI channel inputs from the received noisy sequences. Simulation results validate the performance robustness of PR-NN to realistic magnetic recording channels.

For decoding of convolutional codes, the authors in \cite{Kim18} proposed a neural decoder named N-RSC, which consists of two layers of bi-GRUs each followed by batch normalization units. Simulation results show that the neural decoder N-RSC is robust to variations of the additive white Gaussian noise (AWGN) channel. This neural decoder is also extended to decode Turbo codes. { In \cite{Kim20}, Kim \textit{et al.} constructed feedback codes, called Deepcode, for AWGN channels via deep learning. Deepcode uses an RNN at the encoder side and two-layer bi-GRUs at the decoder side. Simulation results show that the error rate performances of Deepcode can outperform known feedback codes significantly. In \cite{Choukroun22}, Choukroun \textit{et al.} proposed a model-free decoder for error correction codes by using the Transformer architecture. Simulation results on a wide variety of codes demonstrate that Transformer decoder significantly outperforms existing state-of-the-art neural decoders. }

\subsubsection{\textbf{{Model-based approaches}}} {The black-box approaches for channel detection/decoding problems illustrated above are purely data-driven and model-agnostic. They typically require a large amount of data and computing resources. Instead, the model-based deep learning methods combine the existing mathematical algorithms designed for channel detection/decoding and deep learning. They always need a few parameters and can be learned from a limited amount of data and computing resources. According to \cite{Shlezinger23}, the model-based approaches for channel detection/decoding can be classified into two categories: model-aided networks via deep unfolding and structure-oriented deep neural network (DNN) aided inference. }

\textit{\textbf{{Deep unfolding:}}} Deep unfolding is a common method for obtaining deep learning architectures from model-based iterative algorithms \cite{Hershey14}. In particular, it converts an iterative algorithm into a DNN by designing each layer to resemble a single iteration.

In \cite{Samuel19}, the authors proposed a detection network (DetNet) specifically designed for MIMO detection. DetNet is constructed by deep unfolding the iterations of a projected gradient descent algorithm into a DNN. Simulation results show that the method can achieve near optimal detection performance with low computational complexity.

In \cite{Nachmani18}, the authors proposed a neural belief propagation (BP) decoder for linear codes by deep unfolding the iterative BP decoding algorithm into a deep neural network. The neural BP decoder assigns weights to the edges in the Tanner graph and these weights are trained using stochastic gradient descent. By tying the weights of each edges in the Tanner graph at each iteration, the feed-forward architecture of neural BP decoder is transferred into an RNN architecture which is termed BP-RNN. The design of neural BP and BP-RNN are also extended to neural min-sum (MS) decoding in \cite{Nachmani18}. { Subsequently, the neural BP decoder is improved by hyper-graph network \cite{Nachmani19} and active learning \cite{Beery20}.} In \cite{Dai21}, the authors proposed a high-performance neural MS decoding for protograph LDPC codes by fully utilizing the lifting structure of the codes. 

\textit{\textbf{{Structure-oriented methods:}}} { By replacing the channel-dependent part of the detection algorithms with DNNs, structure-oriented DNN-aided methods use deep learning to enhance the robustness of the existing channel detection algorithms. There are a number of structure-oriented detection methods for various communication channels. }

In \cite{Moham19}, the authors proposed a deep learning-based sphere decoding for MIMO detection, where the radius of the decoding hypersphere is learned by a three-layer DNN prior to decoding. Simulation results show that the performance of the method is close to the optimal maximum likelihood decoding with significantly reduced computational complexity.

For finite memory channels such as Poisson channel and ISI channel with AWGN, the authors in \cite{Shlezinger20} proposed a data-driven symbol detector, called ViterbiNet, that does not require CSI. ViterbiNet is designed by replacing the channel-dependent part of Viterbi algorithm with a DNN. The authors also proposed a meta-learning approach to track dynamic channel conditions. Simulation results show that ViterbiNet is robust to CSI uncertainty, i.e., the performance of ViterbiNet without the knowledge of CSI can approach that of the CSI-based Viterbi algorithm. In \cite{Sun21}, Sun \textit{et al.} proposed a non-trivial variation of ViterbiNet based on generative adversarial networks (GAN) for emerging communications systems where the underlying channel models are highly complex or completely unknown. In particular, GAN is used to learn the channel transition probability for the channel-dependent part of Viterbi algorithm. More recently, the DNN used in ViterbiNet is replaced by the concatenation of a convolutional neural network and a bidirectional long short-term memory network \cite{Lan22}.

For BCJR algorithm, the optimal maximum a-posteriori probability (MAP) symbol detector, Shlezinger \textit{et al.} proposed a data-driven detector called BCJRNet \cite{Shlezinger20-2}\cite{Farsad21}, which shows improved robustness to CSI uncertainty when compared with the conventional BCJR detector. Subsequently, the authors in \cite{Tsai20} proposed a BCJRNet receiver for joint symbol detection and channel decoding by using a dedicated neural network to replace the channel-model-based computation in the BCJR receiver. It can achieve 1.0 dB gain when compared with the conventional BCJR algorithm under CSI uncertainty.

The conventional symbol detection methods for ISI channels, such as the Viterbi and BCJR algorithms, are special cases of sum-product algorithm on factor graph \cite{Kschischang01}. In the view of this point, the authors in \cite{Schmid22} developed efficient strategies to improve the performance of the factor graph-based symbol detection based on neural enhancement. In particular, the neural BP and generalizations of factor nodes are used to mitigate the effect of cycles within the factor graph. Simulation results show that the methods can approach MAP performances in various scenarios with low computational complexity.

\subsection{Our Contributions}

Deep learning-based channel detection methods for communication systems with unknown channel models or CSI uncertainty have received a lot of attention in recent years. However, the deep learning-based method has not been explored carefully for channels with insertions and deletions which have potential applications in various communication and storage systems. In this paper, as a first step to developing deep learning-based methods for synchronization channels, we will investigate channel detection methods for marker codes via deep learning. {It is also possible to improve the decoding of marker code via deep learning as in \cite{Choukroun22} or develop deep learning-based joint detection and decoding scheme for maker code as in \cite{Tsai20}. However, these research problems are beyond the scope of this paper and need further studies.} The main contributions of this paper are summarized as follows.

1) We propose a model-based neural network, named FBNet, for symbol detection of marker codes over insertion and deletion channels. FBNet is a custom RNN which is designed by deep unfolding the iterative FB algorithm of marker codes into a feed-forward neural network (FFNN) and then tying the corresponding weights in each layer to make them equal. Sharing weights in each layer transfers the unfolded FFNN into an RNN. These two ideas are used for developing deep learning-based decoding of linear codes, e.g. in \cite{Nachmani18} and \cite{Dai21}, which motivate us to design FBNet. Simulation results show that the detection performance of FBNet with CSI uncertainty can approach that of the CSI-based FB algorithm for marker codes over random synchronization channels.

2) Inspired by the excellent performance of bi-GRU on sequence data \cite{Chung14}, we propose a data-driven neural network based on bi-GRU, named FBGRU, for symbol detection of marker codes. It should be noted that bi-GRU is also used for symbol detection in partial response channels \cite{Zheng21} and channel decoding of convolutional and Turbo codes \cite{Kim18}. The detection performance of FBGRU can approach that of the FB algorithm with perfect CSI. When operating with the same level of CSI uncertainty, the proposed FBGRU can outperform the original FB algorithm for marker codes. Moreover, when the probabilistic channel model of insertions and deletions is unknown and the receiver assumes a known channel model, the proposed FBGRU can outperform both the original FB algorithm and the FBNet. This is due to the fact that FBGRU is data-driven and its performance {does not depend} on the channel model too much.

The remainder of this paper is organized as follows. In Section II, we present the preliminary knowledge of channel models, marker code and the FB algorithm. In Section III, the model-driven neural network FBNet is introduced. Section IV proposes FBGRU, a data-driven neural network based on bi-GRU. Section V details simulation setups and performance comparisons of the proposed deep learning-based detection methods. Finally, Section VI concludes the paper.

\section{Preliminaries}

\subsection{Channel Models Related to Insertions and Deletions}

There exist several channel models related to insertions and deletions in the literature. In this paper, we only consider binary channels. The following channel models are used in subsequent sections.

\subsubsection{IDS Channel}
The IDS channel model is used for watermark code \cite{Davey01} and marker code \cite{Ratzer05}. As can be seen in Fig. 1, the IDS channel is controlled by three parameters, the insertion probability $P_i$, the deletion probability $P_d$, and the substitution probability $P_s$. The binary sequence $\bm{Y}$ is fed into the channel, and a binary sequence $\bm{R}$ is yielded at the channel output. When the $i$th symbol $Y_i$ enters the IDS channel, one of three events takes place: i) $Y_i$ is deleted with probability $P_d$; ii) A single random symbol is inserted with probability $P_i$ with $Y_i$ remaining untransmitted; iii) With probability $P_t = 1-P_d-P_i$, $Y_i$ is transmitted and this transmition suffers from a substitution error with probability $P_s$.

\begin{figure}[tb]\label{Fig1}
	\centering
	\includegraphics[height=0.9in]{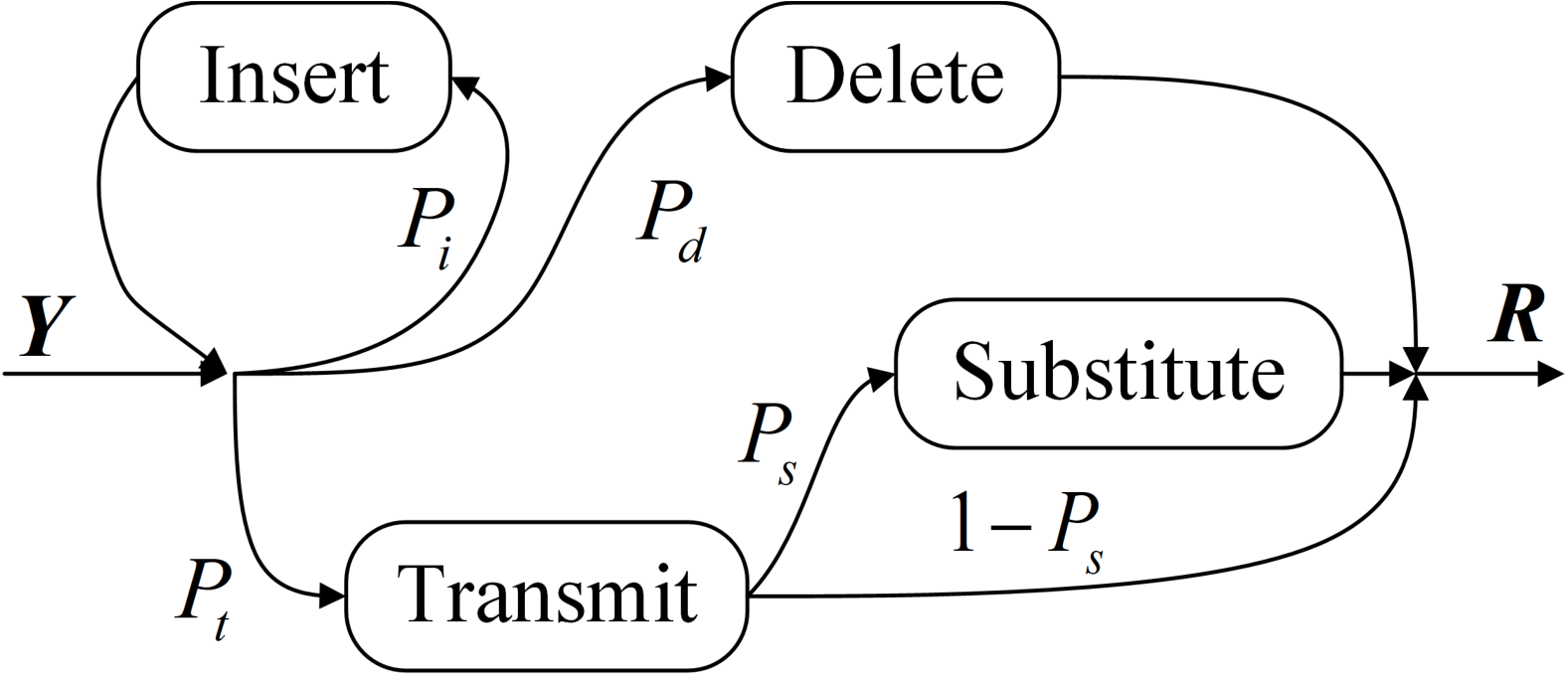}
	\caption{The IDS channel model.}
\end{figure}

\subsubsection{ID-AWGN Channel}
The ID-AWGN Channel model is depicted in Fig. 2. Each transmitted symbol in $\bm{Y}$ is first modulated by binary phase-shift-keying (BPSK), i.e., each symbol $Y_i \in \{0,1\}$ is mapped to $Z_i \in \{-1,+1\}$ such that $Z_i = (-1)^{Y_i}$. Then the modulated symbols go through an IDS channel with $P_s=0$. After that, each symbol output by the IDS channel is corrupted by AWGN with mean $0$ and variance $\sigma^2$. This channel model is investigated in \cite{Jiao12} and \cite{Han13} which is motivated by the bit-patterned media recording system.

\begin{figure}[tb]\label{Fig2}
	\centering
	\includegraphics[height=0.95in]{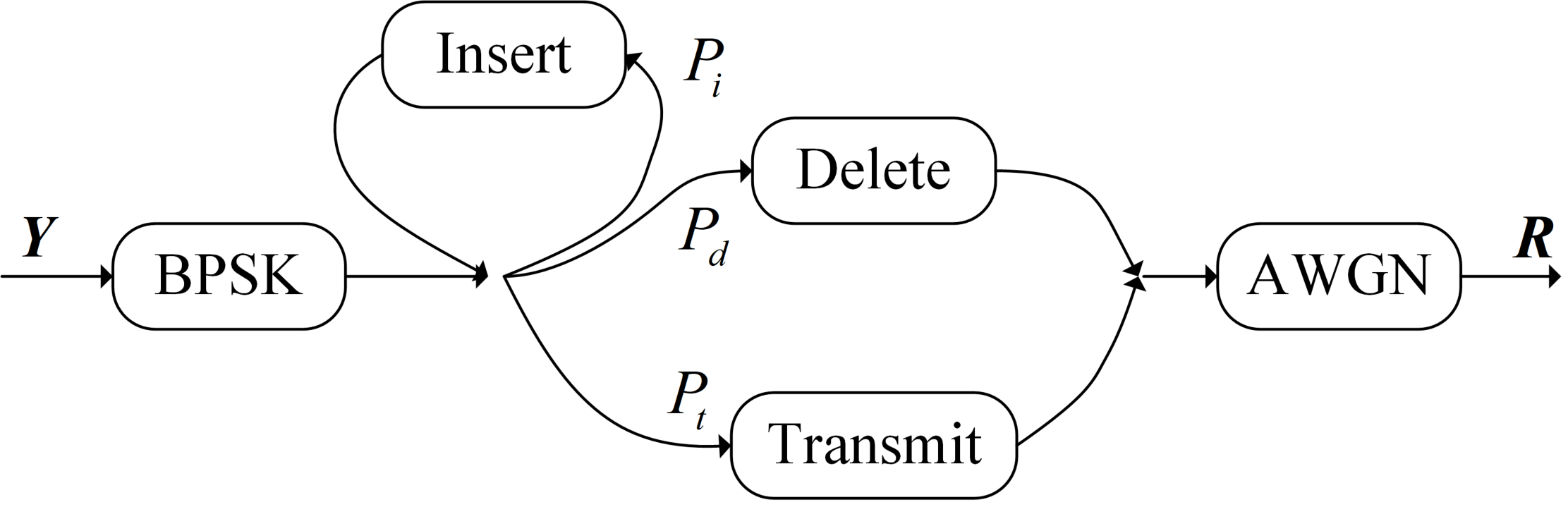}
	\caption{The ID-AWGN channel model.}
\end{figure}

We will devote ourself to design FBNet and FBGRU for ID-AWGN channels in Section III and Section IV, respectively. Then in Section V, the proposed neural networks are adapted for IDS channels by simply changing the input.

\begin{figure*}[t]\label{Fig3}
	\centering	
	\subfigure[]{\includegraphics[width=5.0in]{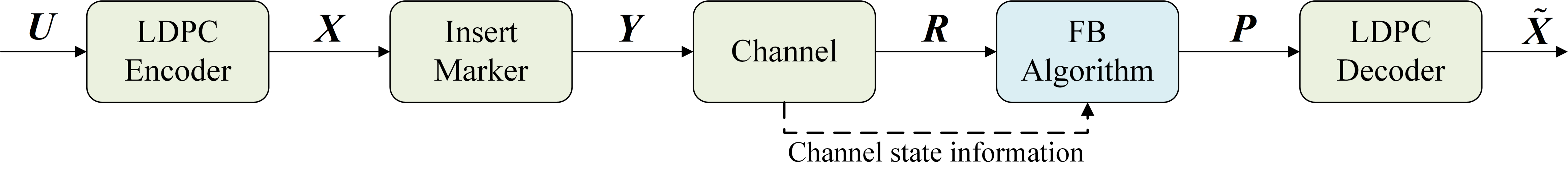}}
	\subfigure[]{\includegraphics[width=5.0in]{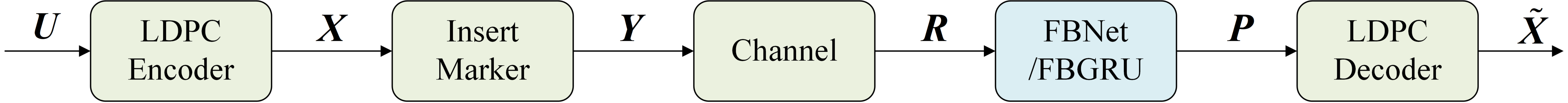}}
	\caption{The system model of marker code: (a) Decoding with the FB algorithm; (b) Decoding with the proposed FBNet/FBGRU.}
\end{figure*}

\subsection{Marker Codes and the Forward-Backward Algorithm}

The system model of marker code for synchronization errors is shown in Fig. 3 (a). A binary information sequence $\bm{U}$ with length $u$ is encoded to $\bm{X}$ with length $x$ by a binary LDPC code. Then a binary marker sequence $\bm{M}$ with length $m$ is inserted into $\bm{X}$ periodically with regular intervals of length $l$. In this paper we choose $\bm{M}=$`001' and then the sequence $\bm{Y}$ to be sent is of length $y=x+3\big(\lceil\frac{x}{l}\rceil-1\big)$. After that, the sequence goes through one of the synchronization channels mentioned in Section II-A and the received sequence $\bm{R}$ of length $r$ is obtained. It should be noted that the length of $\bm{R}$ is a random variable and it may be different from the length of sent sequence. At the decoder side, the log-likelihood ratio (LLR) value of each transmitted bit is inferred by the FB algorithm according to the inserted markers in the send sequence. The LLR values are then fed into LDPC decoder and an estimate $\tilde{\bm{X}}$ of $\bm{X}$ is obtained. In the following, the details of the FB algorithm is presented.

The goal of the FB algorithm is to compute {the a posteriori probabilities} ${\Pr(Y_j|\bm{R}) }$ for $1 \leq j \leq y$. Then the LLR value of the $j$th bit can be calculated as follows:
\begin{IEEEeqnarray}{rll}\label{Eq1}
L(Y_j) = \ln\frac{{\Pr(Y_j=0|\bm{R})}}{{\Pr(Y_j=1|\bm{R})}}.
\end{IEEEeqnarray}
The received vector of synchronization channels can be modeled as being produced by a hidden Markov model (HMM) \cite{Davey01}. Let $S_j$ denote the synchronization \textit{drift} at the $j$th position, which is defined as the number of insertions minus the number of deletions for the first $j$ transmitted bits. The sequence $\bm{S}=\{S_1,\dots,S_y\}$ will form the hidden states of HMM. The synchronization task can be achieved by performing the FB algorithm on a lattice representation of the synchronization \cite{Davey01}\cite{Ratzer05}.

The forward quantity can be defined as
\begin{IEEEeqnarray}{rll}\label{Eq2}
\alpha_j(k)=\Pr(R_1,\dots,R_{j-1+k},S_j=k),  
\end{IEEEeqnarray}
which denotes the probability that the drift at position $j$ is $k$ and the first $j-1+k$ symbols emitted by the channel agree with $\bm{R}$. Similarly, the backward quantity can be defined as
\begin{IEEEeqnarray}{rll}\label{Eq3}
\beta_j(k)=\Pr(R_{j+k},\dots, R_{r}|S_j=k),
\end{IEEEeqnarray}
which denotes the probability of emitting the tail of $\bm{R}$ given a drift of $k$ at position $j$.

For $j\in\{1,2,\dots,y\}$, the forward quantity $\alpha_j(k)$ is computed recursively as
\begin{IEEEeqnarray}{rll}\label{Eq4}
\alpha_j(k)&=&\frac{P_i}{2}\alpha_j(k-1)+P_d\alpha_{j-1}(k+1) \ \ \ \ \ \ \ \  \notag\\
&&+P_t\alpha_{j-1}(k)\sum_{Y_j\in \{0,1\}}\Pr(Y_j)F(Y_j,R_{j+k}),
\end{IEEEeqnarray}
where $\Pr(Y_j)$ is the prior probability of $Y_j$ being $0$ or $1$. When $Y_j$ is an information bit, we assume $\Pr(Y_j=0)=\Pr(Y_j=1)=0.5$. When $Y_j$ is a marker bit, $\Pr(Y_j)=0\ \rm{or}\ 1$ since the marker bits are fixed.
For IDS channel, the metric $F(Y_j,R_{j+k})$ in (\ref{Eq4}) can be calculated as
\begin{IEEEeqnarray}{rll}\label{Eq5}
F(Y_j,R_{j+k})=
\begin{cases}
1-P_s, & \mbox{if }Y_j = R_{j+k} \\
P_s. & \mbox{if }Y_j \neq R_{j+k}
\end{cases}
\end{IEEEeqnarray}
For ID-AWGN channel, we have
\begin{IEEEeqnarray}{rll}\label{Eq6}
F(Y_j,R_{j+k})=
\begin{cases}
\frac{e^{L_{j+k}}}{1+e^{L_{j+k}}}, & \mbox{if }Y_j=0 \\
\frac{1}{1+e^{L_{j+k}}}, & \mbox{if }Y_j=1
\end{cases}
\end{IEEEeqnarray}
where
\begin{IEEEeqnarray}{rll}
L_{j+k}=\ln{\frac{\Pr(Z_{j}=1|R_{j+k})}{\Pr(Z_{j}=-1|R_{j+k})}}=\frac{2R_{j+k}}{\sigma^2}, \notag
\end{IEEEeqnarray}
and $Z_j = (-1)^{Y_j}$ is obtained by BPSK modulating of the symbol $Y_j$ in the ID-AWGN channel. It is evident from Eq. (\ref{Eq5}) and Eq. (\ref{Eq6}) that the equation $F(Y_j=0,R_{j+k})=1-F(Y_j=1,R_{j+k})$ holds.

Similarly, the backward quantity $\beta_j(k)$ is computed recursively as
\begin{IEEEeqnarray}{rll}\label{Eq7}
\beta_j(k)&=&\frac{P_i}{2}\beta_j(k+1)+P_d\beta_{j+1}(k-1)\notag\\
&&+P_t\beta_{j+1}(k)\sum_{Y_{j+1}\in \{0,1\}}\Pr(Y_{j+1})F(Y_{j+1},R_{j+1+k}).
\end{IEEEeqnarray}

Based on the forward and backward quantities, ${\Pr(Y_j|\bm{R})}$ can be calculated as follows
\begin{IEEEeqnarray}{rll}\label{Eq8}
{\Pr(Y_j|\bm{R})}&=&\frac{P_i}{2}\sum_{k=-\delta}^{\delta}{\alpha_{j}(k-1)\beta_{j}(k)}+P_d\sum_{k=-\delta}^{\delta}{\alpha_{j-1}(k+1)\beta_{j}(k)} \notag\\
&&+P_t\sum_{k=-\delta}^{\delta}{\alpha_{j-1}(k)\beta_{j}(k)F(Y_j,R_{j+k})},
\end{IEEEeqnarray}
where $\delta$ denotes the maximum allowed drift and $\alpha_{j}(k) = \beta_{j}(k) = 0$ when $\left\vert k \right\vert > \delta$. The FB algorithm is initialized as follows: For $\alpha_{0}(k)$, we set $\alpha_{0}(0)=1$ and $\alpha_{0}(k)=0$ when $k > 0 $; For $\beta_{y}(k)$, we set $\beta_{y}(r-y)=1$ and $\beta_{y}(k)=0$ when $k < r-y$.

{It should be noted that the equations in FB algorithm described above follow from \cite[Section III]{Wang11}, but there exist some differences. First, the meaning of the notation for the forward quantity is different. In \cite[Section III]{Wang11}, $\alpha_k(n)$ denotes the forward quantity when the system sends $k$ bits and receives $n$ bits. While in Eq. (\ref{Eq4}), $\alpha_j(k)$ denotes the forward quantity when the system sends $j$ bits and receives $j+k$ bits, where $k$ represents the drift. Similar difference also exists in the backward quantity. We change the meaning of these two notations to facilitate the description of FBNet, e.g., when expanding $\alpha_j(k)$ in Eq. (\ref{Eq9}). Second, the terms related to the insertion event in Eqs. (\ref{Eq4}), (\ref{Eq7}), and (\ref{Eq8}) are different from that in \cite[Section III]{Wang11}. This is because the channel model used in our paper and that used in \cite{Wang11} is slightly different. We emphasize that in principle all the FB algorithms described in \cite{Ratzer05}, \cite{Wang11}, \cite{Han13}, and our paper are the same.}

In the following sections, we will design neural network-aided a posteriori probability computing modules FBNet and FBGRU to replace the FB algorithm, as shown in Fig. 3 (b). The proposed methods are more robust to CSI uncertainty or unknown channel models.

\section{FBNet: A Model-Driven Approach}

In this section, the forward and backward computations of the FB algorithm are first unfolded into FFNNs. Then by sharing the weights of each layer, we obtain FBNet\textemdash a self-designed RNN for symbol detection of marker code. Finally, the details of FBNet are presented.

\subsection{Deep Unfolding of the FB Algorithm}
Both the forward and backward computations of the FB algorithm are iterative, and thus can be transformed into neural networks by deep unfolding. In the following, we take the forward quantity calculation for example.

To facilitate the design of the neural network, we first expand $\alpha_j(k-1)$ in the right hand side of (\ref{Eq4}) with the recursive formula. Then $\alpha_j(k)$ can be approximated as follows:

\begin{equation}\label{Eq9}
\begin{aligned}
\alpha_j(k)&=\frac{P_i}{2}\Bigg(\frac{P_i}{2}\alpha_j(k-2)+P_d\alpha_{j-1}(k)+P_t\alpha_{j-1}(k-1)\cdot \\
&\ \ \sum_{Y_{j} \in \{0,1\} }\Pr(Y_{j})F(Y_j,R_{j-1+k})\Bigg) + P_d\alpha_{j-1}(k+1) \\
&\ \ +P_t\alpha_{j-1}(k)\sum_{Y_j \in \{0,1\}}\Pr(Y_j)F(Y_j,R_{j+k})\\
&\overset{{\rm (a)}}{\approx} w_{j1}\alpha_{j-1}(k-1)\sum_{Y_{j} \in \{0,1\} }\Pr(Y_{j})F(Y_j,R_{j-1+k}) \\
&\ \ +w_{j2}\alpha_{j-1}(k)\sum_{Y_j \in \{0,1\} }\Pr(Y_j)F(Y_j,R_{j+k})+w_{j3}\alpha_{j-1}(k+1),
\end{aligned}
\end{equation}
where (a) follows from the following two changes: i) The terms $P_i^2$ and $P_iP_d$ are neglected since the probabilities of two consecutive insertions and one insertion followed by one deletion are quite small; ii) The CSI such as $P_i$, $P_d$, and $P_t$ in (\ref{Eq9}) are replaced by a set of weights $\{w_{j1},w_{j2},w_{j3}\}$ that can be learned through the training algorithm of neural networks. According to (\ref{Eq9}), the computational process of forward quantities can be represented by an FFNN with $y$ layers as shown in Fig. 4. In each layer, there are three types of neurons: input neurons, output neurons, and computing neurons. In general, the input neurons provide necessary input information for the computing neurons, the computing neurons can be designed via (\ref{Eq9}) which will be detailed in Section III-B, and the output neurons provide the forward quantity of each symbol for subsequent calculations. Two consecutive layers are connected via computing neurons. Similarly, the calculation of backward quantities can also be transformed into an FFNN. As the length of the code sequence increases, the number of layers and weights become increasingly large which makes the FFNN difficult to training. Motivated by the RNN decoding of linear codes \cite{Nachmani18}, the weights can be tied, i.e., they are set to be equal in each layer. This tying transfers the FFNN into an RNN which is termed as FBNet. 

\begin{figure}[tb]\label{Fig4}
	\centering
	\includegraphics[width=3.0in]{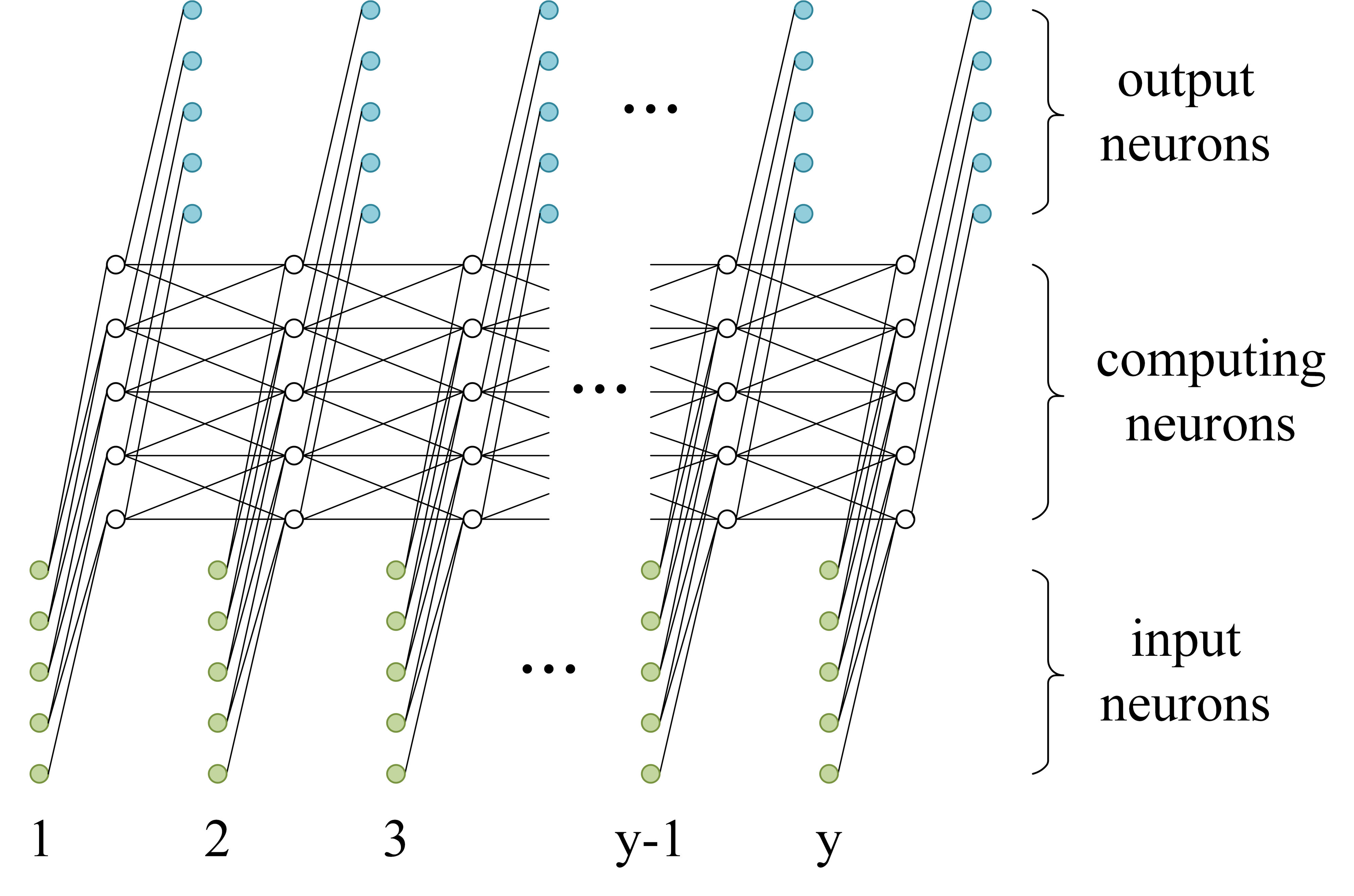}
	\caption{A deep FFNN architecture for calculating the forward quantities.}
\end{figure}

\subsection{The Design of FBNet}

According to the FB algorithm for marker codes, we design FBNet that does not require CSI at the receiver. The structure of FBNet is shown in Fig. 5. It consists of $y$ time steps and in each step the inputs $\{C_j,D_j\}$ are fed into FBNet and $O_j$ is the output. There are three computing units for each time step: the forward cell, the backward cell, and the \textit{a posteriori} probability (APP) unit. To facilitate the design of computing units for FBNet, we first represent the calculation of $\alpha_j(k)$, $\beta_j(k)$, and ${\Pr(Y_j|\bm{R})}$ in the form of vector computations.

\begin{figure}[tb]\label{Fig5}
	\centering
	\includegraphics[width=3.5in]{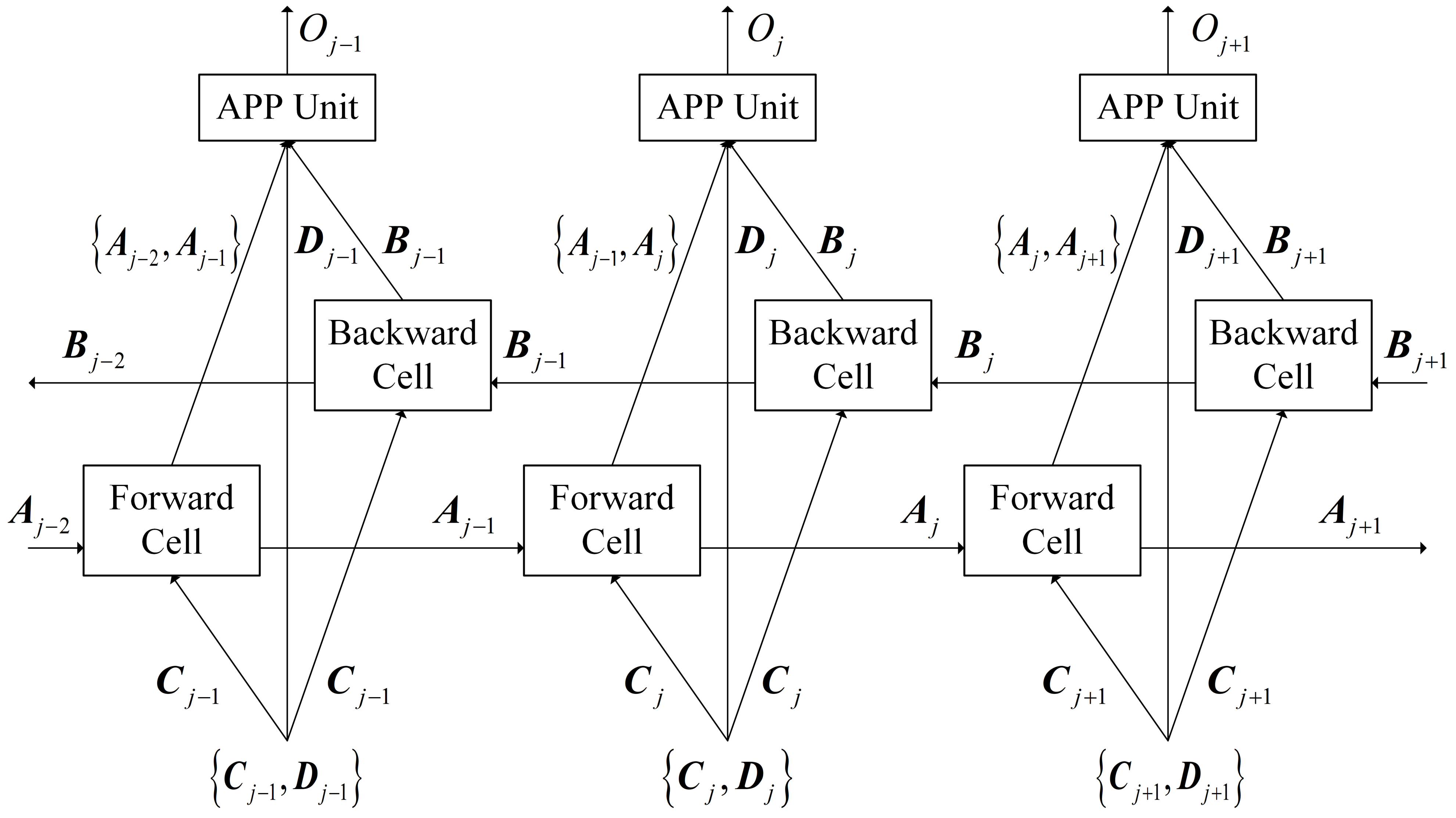}
	\caption{The structure of FBNet.}
\end{figure}

Let $\bm{A}_j$ denote the vector $(\alpha_{j}(-\delta),\dots,\alpha_{j}(0),\dots,\alpha_{j}(\delta))$ of length $2\delta+1$. Let $\overset{\rightharpoonup}{\bm{A}}_j=(0,\alpha_{j}(-\delta),\dots,\alpha_{j}(\delta-1))$ be a vector of length $2\delta+1$ obtained from $\bm{A}_j$ by deleting the {right-most} element in $\bm{A}_j$ and inserting a zero to the left of $\bm{A}_j$. Similarly, we also define a vector $\overset{\leftharpoonup}{\bm{A}}_j=(\alpha_{j}(1-\delta),\dots,\alpha_{j}(\delta),0)$ of length $2\delta+1$ that is obtained from $\bm{A}_j$ by deleting the {left-most} element in $\bm{A}_j$ and appending a zero to the right of $\bm{A}_j$. The introduction of $\overset{\rightharpoonup}{\bm{A}}_j$ and $\overset{\leftharpoonup}{\bm{A}}_j$ is to ensure that the vectors are of the same size in subsequent calculations. The same operations are also used for other vectors if necessary. Let $\bm{\gamma}_j$ denote the vector of length $2\delta+1$ with elements ranging from $\sum_{Y_{j}}\Pr(Y_j)F(Y_j,R_{j-\delta})$ to $\sum_{Y_{j}}\Pr(Y_j)F(Y_j,R_{j+\delta})$. Then the calculation of $\alpha_j(k)$ for {$k\in \{-\delta,\cdots,\delta\}$} can be represented as the following form
\begin{equation}\label{Eq10}
\bm{A}_j=w_1 \overset{\rightharpoonup}{\bm{A}}_{j-1}\odot \overset{\rightharpoonup}{\bm{\gamma}}_j +w_2 \bm{A}_{j-1}\odot \bm{\gamma}_j+w_3 \overset{\leftharpoonup}{\bm{A}}_{j-1},
\end{equation}
where $\odot$ denotes the vector operation of element-wise product. In fact, Eq. (\ref{Eq10}) represents the computation process of a single layer in the FFNN designed for the forward quantity calculation depicted in Fig. 4. The input neurons provide $\bm{\gamma}_j$, while the output neurons output $\bm{A}_j$ which is utilized for subsequent calculations. The weights $\{w_{j1},w_{j2},w_{j3}\}$ in Eq. (\ref{Eq9}) are shared in FBNet and correspondingly represented as $\{w_{1},w_{2},w_{3}\}$ in Eq. (\ref{Eq10}).
Similarly, the calculation of $\beta_{j-1}(k)$ for {$k\in \{-\delta,\cdots,\delta\}$} can be represented as
\begin{equation}\label{Eq11}
\bm{B}_{j-1}=w_5 \overset{\leftharpoonup}{\bm{B}}_{j}\odot\overset{\leftharpoonup}{\bm{\gamma}}_j +w_6 \bm{B}_{j}\odot\bm{\gamma}_j +w_7 \overset{\rightharpoonup}{\bm{B}}_{j},
\end{equation}
where $\bm{B}_{j-1} \triangleq (\beta_{j-1}(-\delta),\dots,\beta_{j-1}(0),\dots,\beta_{j-1}(\delta))$ and the CSI in $\beta_{j-1}(k)$ are replaced with weights $w_5$, $w_6$, and $w_7$.

Finally, based on ${\Pr(Y_j|\bm{R})}$ in Eq. (\ref{Eq8}), FBNet calculates
\begin{equation}\label{Eq12}
\begin{aligned}
	P_j^{1} = w_9 \overset{\rightharpoonup}{\bm{A}}_j {\bm{B}_j}^T+w_{10} \overset{\leftharpoonup}{\bm{A}}_{j-1}{\bm{B}_j}^T +w_{11}\bm{A}_{j-1}\big(\bm{B}_j\odot \bm{\epsilon}_j\big)^T,
\end{aligned}
\end{equation}
and
\begin{equation}\label{Eq13}
\begin{aligned}
P_j^{0} = w_9 \overset{\rightharpoonup}{\bm{A}}_j {\bm{B}_j}^T+w_{10} \overset{\leftharpoonup}{\bm{A}}_{j-1}{\bm{B}_j}^T +w_{12}\bm{A}_{j-1}\big(\bm{B}_j\odot(\bm{1}-\bm{\epsilon}_j)\big)^T,
\end{aligned}
\end{equation}
where $P_j^{1}$ and $P_j^{0}$ denote unnormalized quantities of ${\Pr(Y_j=1|\bm{R})}$ and ${\Pr(Y_j=0|\bm{R})}$, respectively. In Eqs. (\ref{Eq12}) and (\ref{Eq13}), $\bm{\epsilon}_j$ denotes the vector $(F(Y_j=1,R_{j-\delta}),\dots,F(Y_j=1,R_{j+\delta}))$ of length $2\delta+1$ and $\bm{1}-\bm{\epsilon}_j$ denotes the vector $(F(Y_j=0,R_{j-\delta}),\dots,F(Y_j=0,R_{j+\delta}))$ since $F(Y_j=0,R_{j})=1-F(Y_j=1,R_{j})$. The CSI in ${\Pr(Y_j|\bm{R})}$ are replaced with weights $w_9$, $w_{10}$, $w_{11}$, and $w_{12}$.

In the following, we provide the details of each module for FBNet.
\subsubsection{The Construction of Inputs}
According to the vector form representation described above, we can see that the two vectors $\bm{\gamma}_j$ and $\bm{\epsilon}_j$ are required during the calculations. The input vectors $\bm{C}_j$ and $\bm{D}_j$ each with length $2\delta+1$ are constructed to compute $\bm{\gamma}_j$ and $\bm{\epsilon}_j$, respectively.

Based on Eq. (\ref{Eq6}), each element of $\bm{\gamma}_j$ can be computed as
\begin{equation}\label{Eq14}
\begin{aligned}
\sum_{Y_{j}\in \{0,1\}}\Pr(Y_j)F(Y_j,R_{j})&=\Pr(Y_j=1)\frac{1}{1+e^{\frac{2}{\sigma^2}R_j}} \\
&\ \ +\Pr(Y_j=0)\frac{1}{1+e^{-\frac{2}{\sigma^2}R_j}},
\end{aligned}
\end{equation}
where $\sigma^2$ is the noise variance of AWGN channel. To make FBNet {CSI-agnostic}, the term $2/\sigma^2$ in (\ref{Eq14}) is replaced by a weight $w$ which can be learned by the training algorithm. We also introduce the sigmoid function
\begin{equation}\label{Eq15}
f_{\rm sigmoid}(p)=\frac{1}{1+e^{-p}}.
\end{equation}
Then we have $F(Y_j=1,R_j)=f_{\rm sigmoid}(w R_j)$ and $F(Y_j=0,R_j)=f_{\rm sigmoid}(-w R_j)$. Next we construct the input $\bm{C}_j$ to make $\bm{\gamma}_j=f_{\rm sigmoid}(w \bm{C}_j)$ for $j=1,2,\cdots,y$. The construction accords to the value of $Y_j$ and the condition whether $Y_j$ is an information bit or a marker bit.
\begin{itemize}
\item When $Y_j=1$ and it is a marker bit, we have $\Pr(Y_j=1)=1$ and $\Pr(Y_j=0)=0$. Then $\bm{\gamma}_j=\big(F(Y_j=1,R_{j-\delta}),\dots,F(Y_j=1,R_{j+\delta})\big)$. We can construct $\bm{C}_j=(R_{j-\delta},\dots,R_{j},\dots,R_{j+\delta})$ to make $\bm{\gamma}_j=f_{\rm sigmoid}(w \bm{C}_j)$.
\item When $Y_j=0$ and it is a marker bit, we have $\Pr(Y_j=1)=0$ and $\Pr(Y_j=0)=1$. Then $\bm{\gamma}_j=\big(F(Y_j=0,R_{j-\delta}),\dots,F(Y_j=0,R_{j+\delta})\big)$. Since $F(Y_j=0,R_j)=F(Y_j=1,-R_j)$, we can construct $\bm{C}_j=(-R_{j-\delta},\dots,-R_{j},\dots,-R_{j+\delta})$ to make $\bm{\gamma}_j=f_{\rm sigmoid}(w \bm{C}_j)$.
\item When $Y_j$ is an information bit and it is assumed that $\Pr(Y_j=0)=\Pr(Y_j=1)=0.5$, we have $\sum_{Y_{j}\in \{0,1\}}\Pr(Y_j)F(Y_j,R_{j})=0.5(F(Y_j=0,R_j)+F(Y_j=1,R_j))=0.5$. It can be seen that all elements of $\bm{\gamma}_j$ is 0.5. In this case, we can construct $\bm{C}_j$ as all zero vector to make $\bm{\gamma}_j=f_{\rm sigmoid}(w \bm{C}_j)=\bm{0.5}$.
\end{itemize}

Similarly, we can construct $\bm{D}_j=(R_{j-\delta},\dots,R_{j},\dots,R_{j+\delta})$ to make $\bm{\epsilon}_j=\big(F(Y_j=1,R_{j-\delta}),\dots,F(Y_j=1,R_{j+\delta})\big) = f_{\rm sigmoid}(w \bm{D}_j)$.

\begin{figure}[tb]\label{Fig6}
	\centering	
	\includegraphics[width=2.8in]{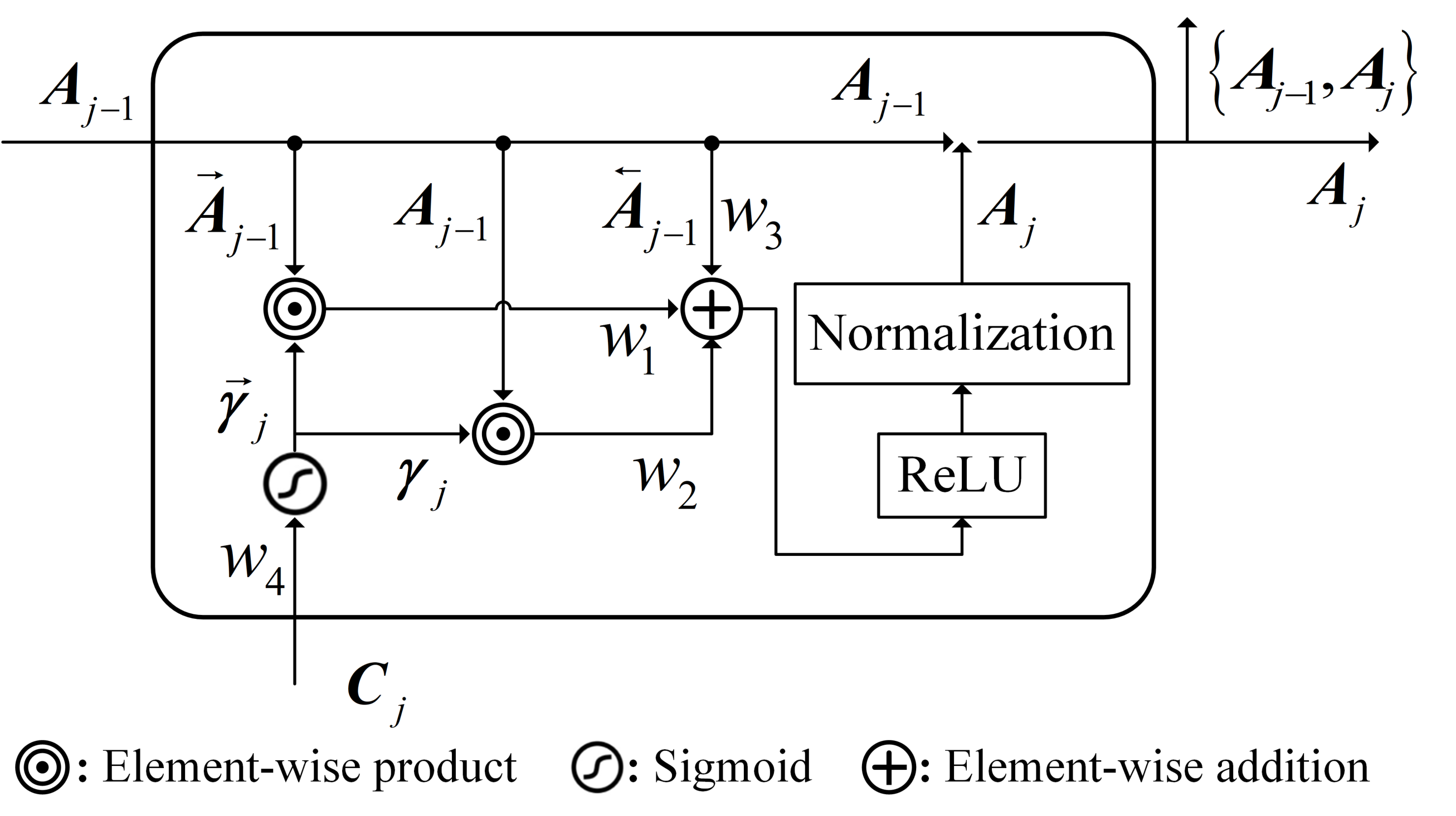}
	\caption{Computing structure of the $j$th forward cell in FBNet.}
\end{figure}

\subsubsection{The Forward Cell}
The inputs to the forward cell are the input vector $\bm{C}_j$ and $\bm{A}_{j-1}$ that comes from the previous time step. The goal of the forward cell is to compute $\bm{A}_{j}$ from $\bm{A}_{j-1}$ and $\bm{C}_j$. Fig. 6 shows the computing structure of the forward cell. By introducing the weight $w_4$, we can calculate $\bm{\gamma}_j=f_{\rm sigmoid}(w_4 \bm{C}_j)$. Then $\bm{A}_{j}$ is obtained according to Eq. (\ref{Eq10}). Since the elements in $\bm{A}_{j}$ are all probabilities, we use a rectified linear units (ReLU) activation function $f_{\rm ReLU}(p)={\rm max}\{0,p\}$ followed by a normalization step to ensure these elements are all confined to $[0,1] \subset \mathbb{R}$. Finally, the calculated vector $\bm{A}_{j}$ is fed into the ($j+1$)th time step and the concatenated vectors $\{\bm{A}_{j-1}, \bm{A}_{j} \}$ are fed into the APP unit. At the first time step, the vector $\bm{A}_{0}$ is initialized as ($0,\dots,1,\dots,0$), in which the middle element is set to 1 and the remaining elements are set to 0.

\subsubsection{The Backward Cell}
Similarly, the computing structure of the backward cell is depicted in Fig. 7. The input vector $\bm{C}_j$ is multiplied by a weight $w_8$, and then $\bm{\gamma}_j$ is calculated as $f_{\rm sigmoid}(w_8 \bm{C}_j)$. The output vector $\bm{B}_{j-1}$ is obtained by using Eq. (\ref{Eq11}), $f_{\rm ReLU}$ and the normalization step. At the last time step, for the vector $\bm{B}_{y}$ of length $2\delta+1$, we initialize the ($r-y+\delta+1$)th element to 1 and the remaining elements to 0.

\begin{figure}[tb]\label{Fig7}
	\centering	
	\includegraphics[width=2.8in]{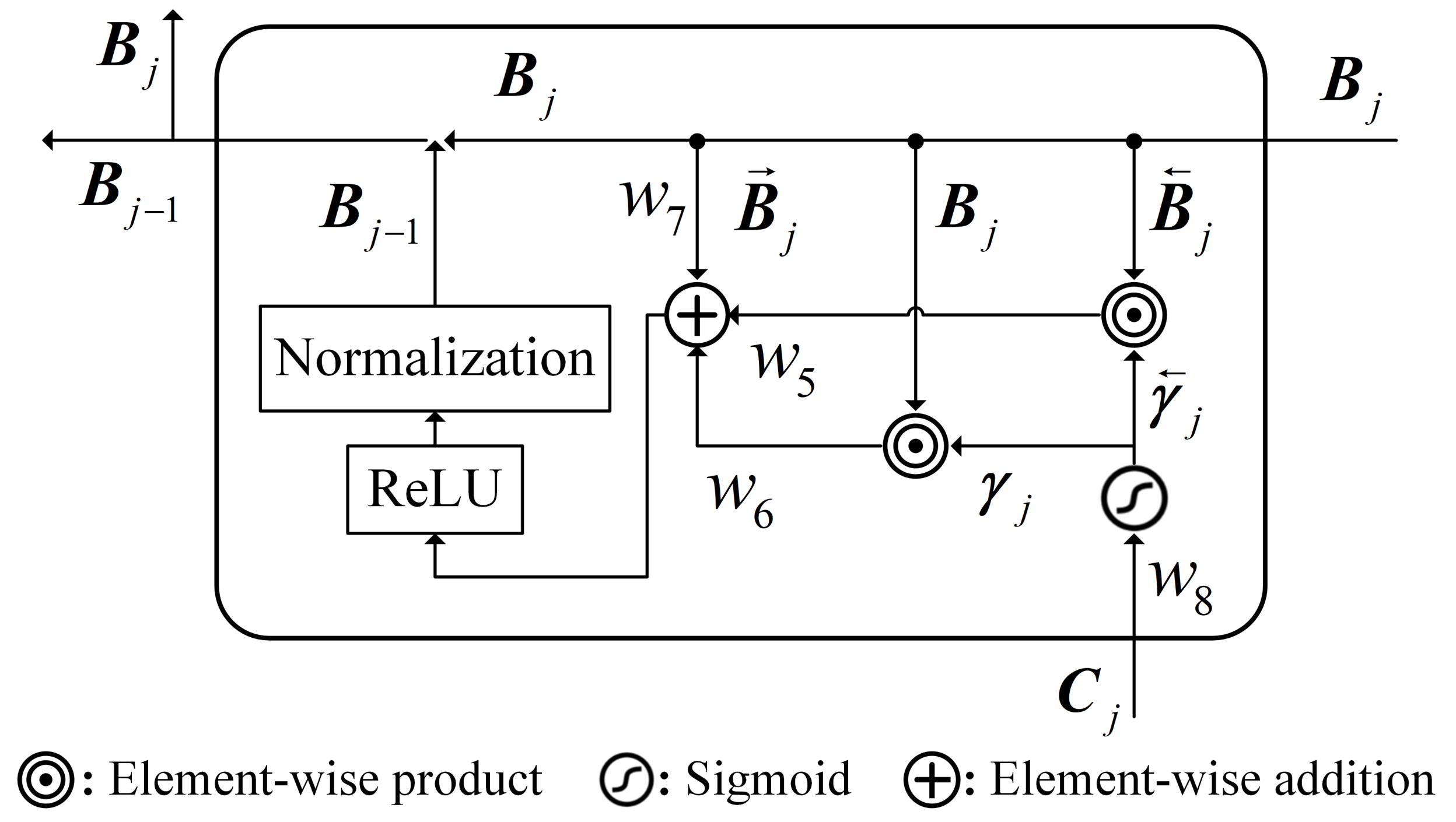}
	\caption{Computing structure of the $(j-1)$th backward cell in FBNet.}
\end{figure}

\subsubsection{The APP Unit}
The goal of the APP unit is to compute the normalized probability

\begin{equation}\label{Eq16}
\begin{aligned}
O_j &= {\Pr(Y_j=1|\bm{R})} = \frac{P_j^1}{P_j^0+P_j^1} = \frac{1}{1+ e^{\ln (P_j^0 / P_j^1 ) }} \\
&= \frac{1}{1+ e^{- \ln (P_j^1 / P_j^0 ) }} =  f_{\rm sigmoid}\Big( \ln P_j^{1} - \ln P_j^{0}\Big).
\end{aligned}
\end{equation}
To avoid possible numeric overflows, we use sigmoid function to calculate $O_j$ instead of direct computation in the above equation. The computing structure of the APP unit is shown in Fig. 8. First, the input vector $\bm{D}_j$ is multiplied by a weight $w_{13}$ and the function $f_{\rm sigmoid}$ is applied to obtain $\bm{\epsilon}_j$. Then $P_j^{1}$ and $P_j^{0}$ are computed according to Eqs. (\ref{Eq12}) and (\ref{Eq13}). To avoid numeric overflow when applying $\ln(\cdot)$ function, the two values $P_j^{1}$ and $P_j^{0}$ should not be too close to 0. To this end, we use the following nonlinear activation function
\begin{IEEEeqnarray}{rll}\label{Eq20}
f_{\rm HardTanh}(p)=
	\begin{cases}
		10^{-12}, & \mbox{if }p < 10^{-12} \\
		1, & \mbox{if } p > 1 \\
		p. & \mbox{others}
	\end{cases} \notag
\end{IEEEeqnarray}
to limit the two values lie in the range $\big[10^{-12}, 1\big] \subset \mathbb{R}$. Finally, the value $\ln P_j^{1}-\ln P_j^{0}$ is computed and then the function $f_{\rm sigmoid}$ is applied to obtain $O_j$.

\begin{figure}[tb]\label{Fig8}
	\centering	
	\includegraphics[width=2.3in]{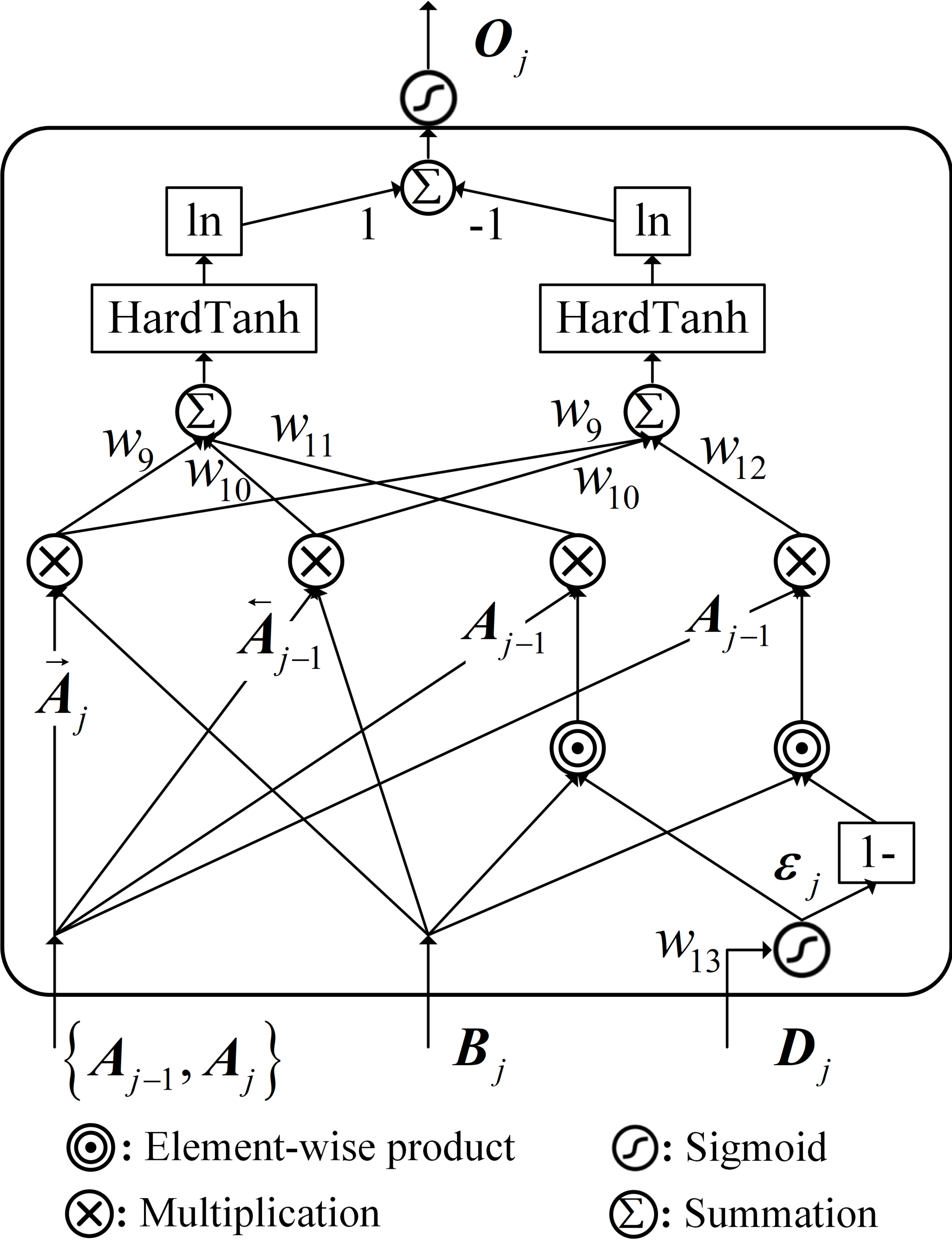}
	\caption{Computing structure of the $j$th APP unit in FBNet.}
\end{figure}

It can be seen that there are 13 weight parameters $w_1 \sim w_{13}$ need to be learned in FBNet. During the training of FBNet, we choose the sequence $\bm{Y}$ as labels, and the loss function we used is the following binary cross entropy function
\begin{equation} \label{Eq17}
f_{\rm{BCE}}(\bm{Y},\bm{O})=-\frac{1}{y}\sum_{j=1}^{y}\Big(Y_j\log{O_j}+(1-Y_j)\log(1-O_j)\Big),
\end{equation}
where $\bm{O}=\{O_1,O_2,\dots,O_y\}$ are the outputs of FBNet.

It should be noted that FBNet is constructed based on the FB algorithm. Therefore, the computational complexity of FBNet is similar to that of the original FB algorithm. Moreover, in  our design of FBNet, the events of two consecutive insertions/deletions and one insertion followed by one deletion are neglected. These events can also be considered in the design of FBNet. This will lead to more complicated computing structures of forward and backward cells.

{It is also possible to design a bidirectional RNN with forward and backward cells for the detection of marker codes by directly using Eq. (\ref{Eq8}) and the loss function (\ref{Eq17}). However, there exist two drawbacks for this bidirectional RNN: 1) The calculation of $\Pr(Y_j|\bm{R})$ in Eq. (\ref{Eq8}) depends on the recursive computations of the forward quantity $\alpha_j(k)$ in Eq. (\ref{Eq4}) and the backward quantity $\beta_j(k)$ in Eq. (\ref{Eq7}). The computational dependency in these two quantities makes the computations in each forward or backward cell difficult to parallelization. 2) Due to the recursive computations in Eqs. (\ref{Eq4}) and (\ref{Eq7}), each forward or backward cell will have multiple neural network layers. The number of layers in each cell is $2\delta+1$ where $\delta$ is the maximum allowed drift. This makes the whole bidirectional RNN deep and leads to gradient explosion or dispersion problems during training. To avoid the above problems, we design FBNet by using approximation on the computations of forward and backward quantities. This reduces the computational dependency and makes the computations in each cell parallel via matrix-vector product.}

{ At the end of this section, we highlight the differences between FBNet and the related work: }

{1) FBNet vs. BP-RNN \cite{Nachmani18}: Though the high-level design of FBNet is similar to BP-RNN decoding of linear codes, the design details are different. The BP-RNN decoding is based on the structure of Tanner graph of linear codes and is performed over memoryless channels. While the design of FBNet is based on the FB algorithm, and the insertion/deletion channel has memory.}

{2) FBNet vs. BCJRNet \cite{Shlezinger20-2}: First, it should be noted that BCJR algorithm is essentially the same as the FB algorithm used for marker codes. However, the communication channels for BCJRNet and FBNet are different. The memory length of ISI channel used for BCJRNet is fixed. While the memory length of insertion/deletion channel used for FBNet is variable and it depends on the positions of insertion/deletion errors. This makes the design of FBNet more complex when compared with BCJRNet. Moreover, BCJRNet is designed by replacing the part of channel-dependent computation in BCJR algorithm with DNN. Thus, BCJRNet is a structure-oriented DNN-aided inference method. While FBNet is derived by unrolling the FB algorithm and it is a model-aided network via deep unfolding.}

\section{FBGRU: A Data-Driven Approach}

In {the previous section}, we propose a model-driven approach FBNet that is heavily relied on the FB algorithm for the detection of marker codes. The proposed FBNet is {CSI-agnostic} and has a few number of parameters which can be trained easily with a small data set. However, the performance of FBNet may be degraded significantly when the channel model is unknown and different from the IDS and ID-AWGN channels wherein the insertion and deletion events are independent with each symbol. For example, we can consider a more bursty synchronization channel in which the consecutive symbol insertions and deletions with small length {occur}. It should be noted that both FBNet and FBGRU are RNN-based, but the computing unit is different. For FBNet, the computing unit is newly designed based on the FB algorithm. While for FBGRU, we use a standard GRU that is frequently used for RNN-based deep learning methods.

\subsection{The Gating Mechanisms for RNN}
RNNs are made of recurrent units used to capture evolution in time, e.g., sequences, time series. When it was first proposed, the nonlinear activation function
\begin{equation}
f_{\rm tanh}(p)=\frac{e^p-e^{-p}}{e^p+e^{-p}}
\end{equation}
is often used in the recurrent unit. However, it is difficult to train RNNs to capture long-term dependencies due to the {effects} of gradient explosion and gradient dispersion. Moreover, as the sequence length increases, the effect of long-term dependencies is hidden by the effect of short-term dependencies. To address this issue, Hochreiter \textit{et al.} proposed a gated RNN, named Long Short-Term Memory (LSTM) network model, by introducing three gates, forgetting gates, input gates, and output gates in LSTM, and memory cells (a special hidden state) to improve the gradient disappearance and gradient explosion of traditional RNN. Later, Cho \textit{et al.} proposed a simple gating mechanism, named GRU network model, to avoid the same problem that LSTM does. Empirical evaluation shows that the performance of GRU and LSTM are similar, but GRU always has fewer parameters and faster convergence rate when compared with LSTM.

{Many problems in natural language processing (NLP) can be modeled as hidden Markov models (HMMs). RNNs have been utilized successfully in the field of NLP involving time sequential data modeled by HMM. As described in Section II, the received vector of IDS/ID-AWGN channels can be modeled as being produced by a HMM. Therefore, RNN-based architecture is a natural candidate for the symbol detection problem of maker codes. Since gated recurrent unit (GRU) is a frequently used RNN cell, the use of GRU for symbol detection of marker codes is feasible and worth exploring.} The computing structure of GRU is shown in Fig. 9. The detailed calculations in a GRU can be formulated as follows
\begin{equation}
\begin{aligned}
	\bm{z}_j&=f_{\rm sigmoid}\big(\bm{W}_z\bm{x}_j + \bm{U}_z\bm{h}_{j-1} + \bm{b}_z \big), \notag \\
    \bm{r}_j&=f_{\rm sigmoid}\big(\bm{W}_r\bm{x}_j + \bm{U}_r\bm{h}_{j-1} + \bm{b}_r \big), \notag \\
    \tilde{\bm{h}}_j&=f_{\rm tanh}\big(\bm{W}_{\tilde{h}}\bm{x}_j + \bm{U}_{\tilde{h}} {\bm{r}_j\odot\bm{h}_{j-1}} + \bm{b}_{\tilde{h}} \big), \notag \\
    \bm{h}_j&=(1-\bm{z}_j)\odot\bm{h}_{j-1}+\bm{z}_j\odot\tilde{\bm{h}}_{j}, \notag
\end{aligned}
\end{equation}
where $\bm{W}_z$, $\bm{U}_z$, $\bm{W}_r$, $\bm{U}_r$, $\bm{W}_{\tilde{h}}$, and $\bm{U}_{\tilde{h}}$ are weight matrices, $\bm{b}_z$, $\bm{b}_r$, and $\bm{b}_{\tilde{h}}$ are biases. The activation function used in a GRU are $f_{\rm sigmoid}$ and $f_{\rm tanh}$. The vectors $\bm{x}_j$ and $\bm{h}_{j-1}$ denote, respectively, the input vector of the $j$th time step and the hidden state output by the GRU computing cell in the $(j-1)$th time step. The vector $\bm{h}_{j}$ is the hidden state output by the $j$th time step. The vectors $\bm{z}_j$ and $\bm{r}_j$ represent the reset gate and the update gate, respectively. The operation $\odot$ denotes the element-wise product of matrices.

\begin{figure}[tb]\label{Fig9}
	\centering
	\includegraphics[width=2.4in]{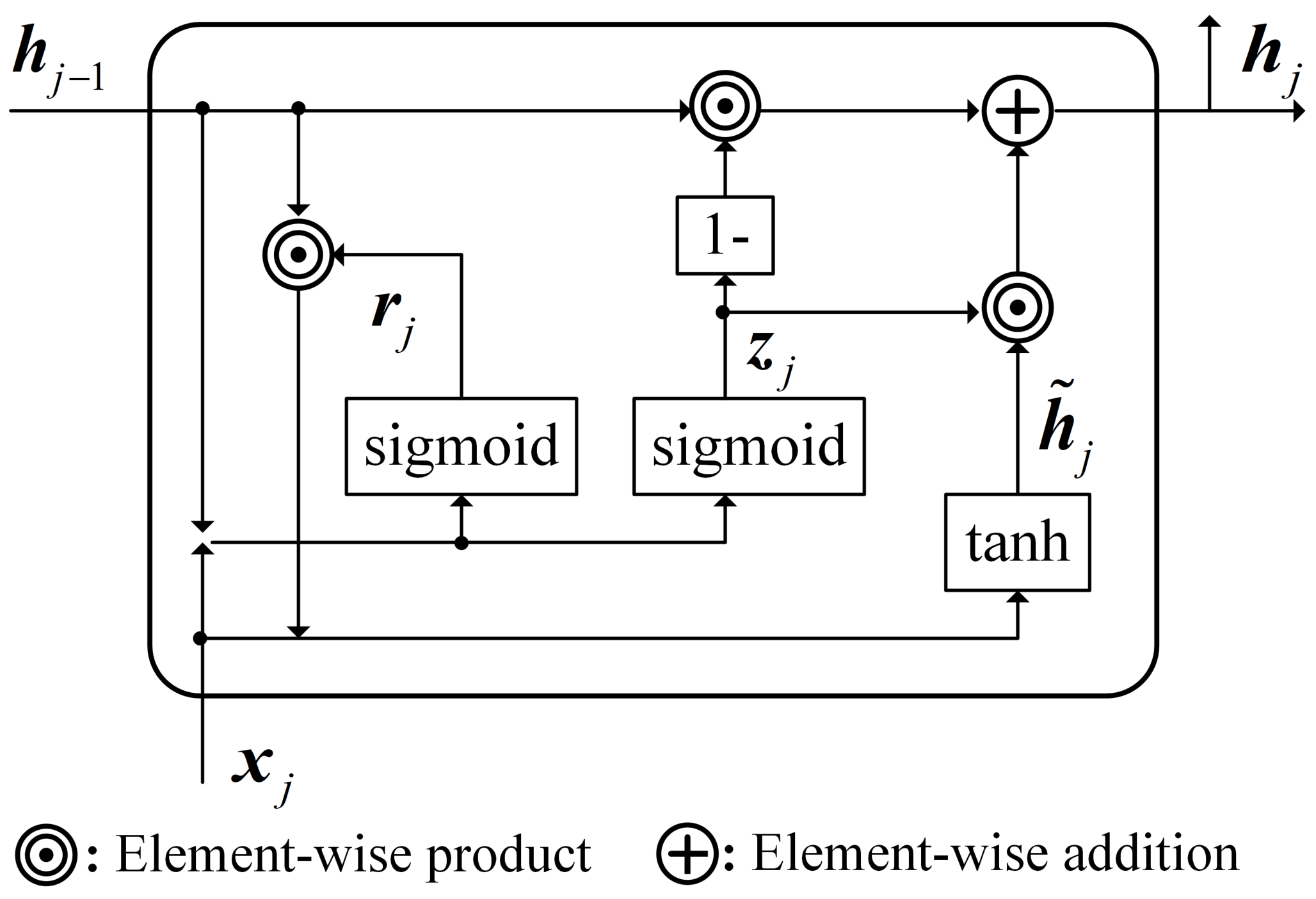}
	\caption{Computing structure of a GRU cell.}
\end{figure}

\subsection{The Design of FBGRU}
The proposed FBNet in the previous section uses forward and backward calculations to infer synchronization. {It should be noted that FB algorithm is designed for random and independent insertion/deletion error events. By deep unfolding the FB algorithm and sharing weights in each layer, FBNet becomes a bidirectional RNN architecture. The computations of forward and backward cells are designed based on the original FB algorithm. If we replace the computation units in FBNet by some universal computation cells from neural networks, the reliance on the condition of random and independent error events in the detection algorithm may be reduced. This motivates us to design FBGRU, which uses GRU and fully connected neural network to replace the forward/backward cells and the APP unit in FBNet, respectively.}

Motivated by the use of bi-direction computations in the design of FBNet, and the use of bi-GRU for the decoding of convolutional codes \cite{Kim18} and the detecting of coded partial-response channels \cite{Zheng21}, we use {bi-GRU cells to design FBGRU}. For each time step, the bi-GRU consists of two separate GRU cells, one operating in the forward direction and the other operating in the backward direction, as shown in Fig. 10. Bi-GRU is a special GRU architecture that has the capability to capture dependence in both forward and backward directions in time. The outputs of a bi-GRU cell in forward and backward directions are merged at each time step.

\begin{figure}[tb]\label{Fig10}
	\centering
	\includegraphics[width=1.45in]{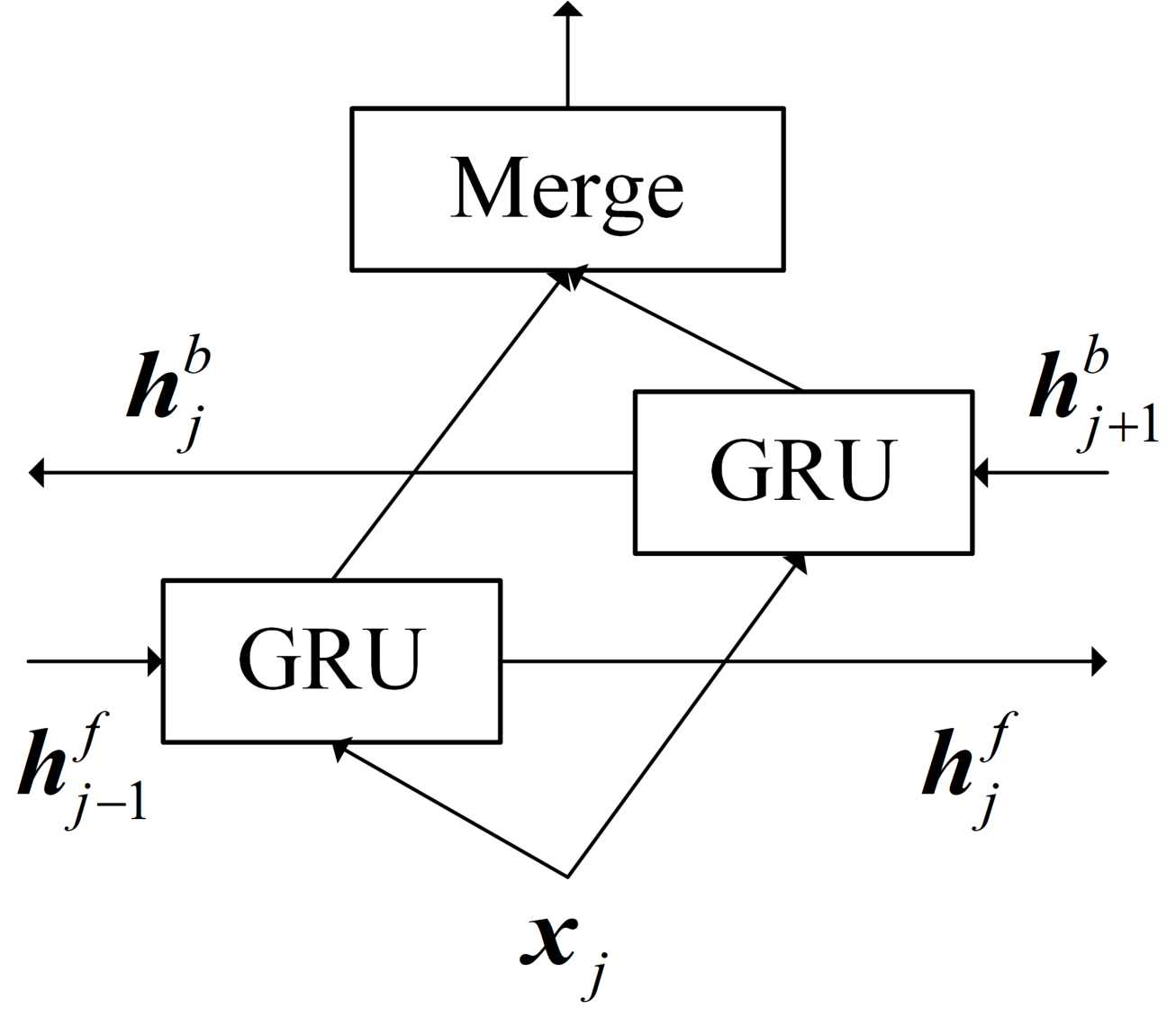}
	\caption{The structure of a bi-GRU cell.}
\end{figure}

The structure of the proposed FBGRU is depicted in Fig. 11. There are $y$ time steps for FBGRU, which is the same as the sequence length. In the following, we present the details of each module in FBGRU.

\begin{figure}[tb]\label{Fig11}
	\centering
	\includegraphics[width=3.0in]{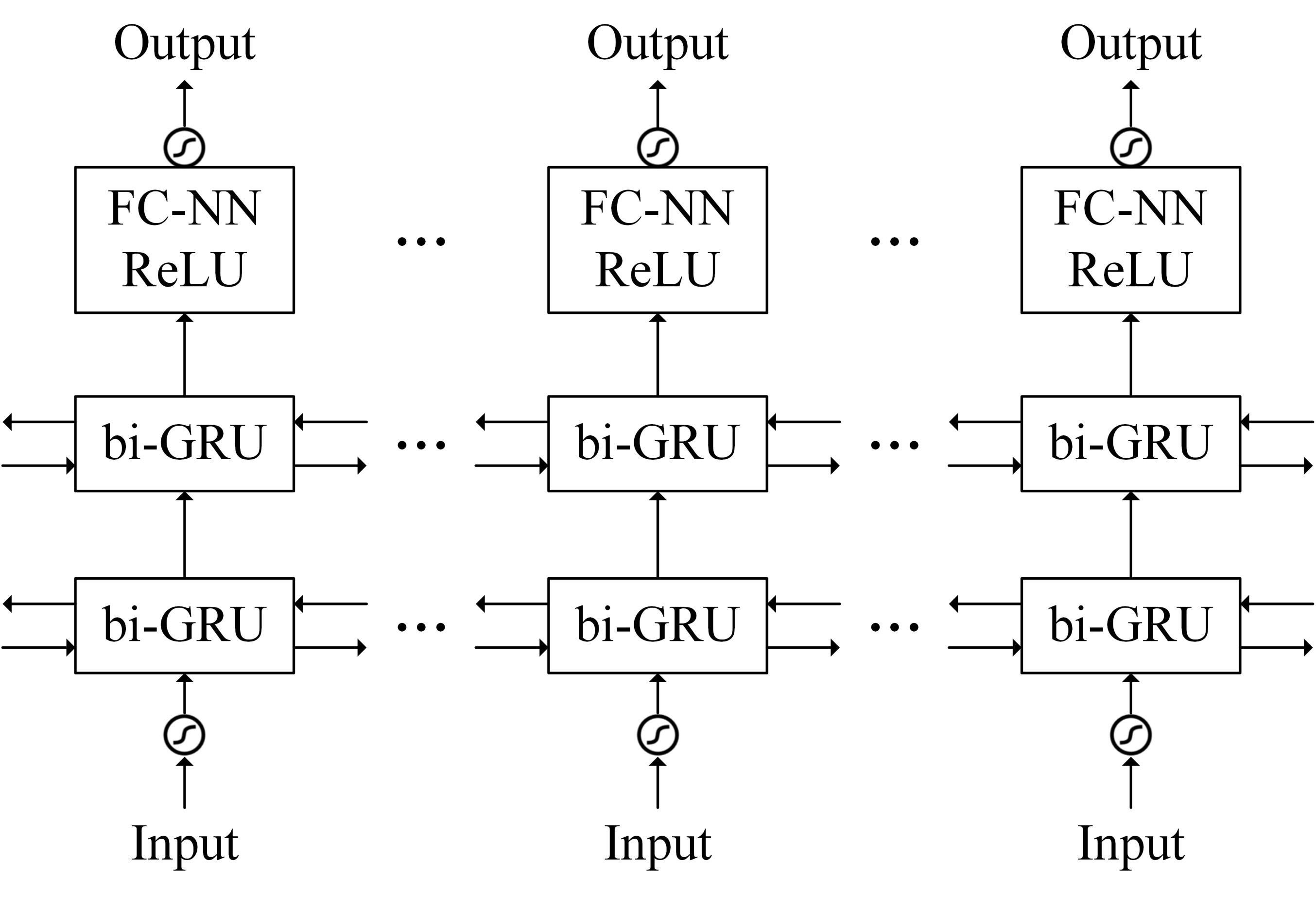}
	\caption{The structure of the proposed FBGRU.}
\end{figure}

\subsubsection{The Input of FBGRU}
Intuitively, using the received sequence $\bm{R}$ as the input of FBGRU is not reasonable since the length of $\bm{R}$ is not fixed due to insertions and deletions. In Section III-B, the input vectors $\bm{C}_j$ and $\bm{D}_j$ are constructed for each time step of FBNet. This construction is based on the MAP rule of the FB algorithm. Therefore, it is convincible that $\bm{C}_j$ and $\bm{D}_j$ contain enough feature information for synchronizing the sequence. For FBGRU, we also choose $\bm{C}_j$ and $\bm{D}_j$ as input vectors for the $j$th time step. Then we introduce weights $w_c$ and $w_d$, and compute $f_{\rm sigmoid}(w_c\bm{C}_j)$ and $f_{\rm sigmoid}(w_d\bm{D}_j)$ as the inputs of the bi-GRU module.

\subsubsection{The Bi-GRU Layers}
The proposed FBGRU contains two bi-GRU layers. For the $j$th time step in the first bi-GRU layer, the forward GRU cell calculates the forward hidden state $\bm{h}_j^{f1}$ based on the input vectors and the forward hidden state $\bm{h}_{j-1}^{f1}$ from the $(j-1)$th time step, and the backward GRU cell calculates the backward hidden state $\bm{h}_{j}^{b1}$ based on the input vectors and the backward hidden state $\bm{h}_{j+1}^{b1}$ from the $(j+1)$th time step. In the second bi-GRU layer, the $j$th bi-GRU cell takes the hidden states $\bm{h}_j^{f1}$ and $\bm{h}_{j}^{b1}$ as input vectors and outputs the hidden states $\bm{h}_j^{f2}$ and $\bm{h}_{j}^{b2}$. The bi-GRU cells in the second layer perform similar calculations as that in the first layer.
We set the default initial hidden states of bi-GRU cells to $\bm{0}$ in both forward and backward directions. {It should be noted that the number of bi-GRU layers in FBGRU can be set as other values. We conduct simulations for FBGRU with one layer, two layers, and three layers of bi-GRUs, respectively. Simulation results show that the error rate performances are improved gradually as the number of layers increases from one to three. }

\subsubsection{The Fully Connected Layers}
The output vectors $\bm{h}_j^{f2}$ and $\bm{h}_{j}^{b2}$ from the bi-GRU layers are fed into a fully connected neural network (FC-NN) block. The designed FC-NN has four fully-connected layers. For the first three layers, we use $f_{\rm ReLU}$ as the activation function, which can result in a nonlinear system and augment the learning capability of the neural network. For the last layer, we use $f_{\rm sigmoid}$ as the activation function to ensure the output $O_j$ is a probability value that lies in the range $[0,1] \subset \mathbb{R}$.

The loss function we used for training FBGRU is the binary cross entropy function $f_{\rm{BCE}}(\bm{Y},\bm{O})$ in (\ref{Eq17}) that is the same as that for FBNet.

\section{Experimental Results}

In the above two sections, the proposed FBNet and FBGRU are designed for ID-AWGN channels. To adapt them for IDS channels, we first map each symbol $R_j \in \{0, 1\}$ in the received sequence $\bm{R}$ to $\theta_j \in \{-1, +1\}$ by the mapping $0 \rightarrow +1$ and $1 \rightarrow -1$. {The} metric $F(Y_j,R_{j+k})$ calculated in (\ref{Eq5}) can be rewritten as
\begin{IEEEeqnarray}{rll} \label{Eq18}
F(Y_j,R_{j+k})=
\begin{cases}
1-P_s, & \mbox{if }Y_j = 0, \theta_{j+k}=+1  \\
P_s, & \mbox{if }Y_j = 0,  \theta_{j+k}=-1  \\
P_s, & \mbox{if }Y_j = 1, \theta_{j+k}=+1  \\
1-P_s. & \mbox{if }Y_j = 1,  \theta_{j+k}=-1  \\
\end{cases}
\end{IEEEeqnarray}
Since $P_s$ is a part of the CSI that may be unknown by the receiver, we can use the approximation $P_s \approx \frac{1}{1+e^w}$ for (\ref{Eq18}) and then $F(Y_j,R_{j+k})$ can be further rewritten as
\begin{IEEEeqnarray}{rll} \label{Eq19}
F(Y_j,R_{j+k})\approx
\begin{cases}
\frac{e^{w\theta_{j+k}}}{1+e^{w\theta_{j+k}}}, & \mbox{if }Y_j=0, \\
\frac{1}{1+e^{w\theta_{j+k}}}, & \mbox{if }Y_j=1.
\end{cases}
\end{IEEEeqnarray}
where $w$ is a weight that can be learned by the training algorithm. After this transformation, the calculation of $F(Y_j,R_{j+k})$ is similar to that for ID-AWGN channels and the designed FBNet and FBGRU can be applied to IDS channels.

\subsection{Experiment Settings}
First, {to validate the error rate performances of FBNet and FBGRU}, we choose two binary LDPC codes {with different structures and rates} for our experiments: 1) A (273,191) {random} LDPC code \cite{MacKay01} $\mathcal{C}_1$ with code length $N=273$, message length $K=191$, and rate $R=0.70$; 2) A (648,540) {standardized quasi-cyclic} LDPC code \cite{RPTU} $\mathcal{C}_2$ with $N=648$, $K=540$, and $R=0.83$. For LDPC decoding, we use the well-known sum-product algorithm with the maximum number of iterations being 30. For all simulations, the marker sequence $\bm{M}=$`001' is inserted into LDPC codeword sequence between every $9$ bits.

Second, according to \cite{Davey01}, the maximum allowed drift $\delta$ is chosen to be several times larger than the standard deviation of the synchronization drift over one block length, given by $\sqrt{y(P_d)/(1-P_d)}$. For convenience, we choose $\delta=17$ for all experiments, which is large enough for the simulated scenarios.

Based on the above parameters, the number of time steps for the proposed neural networks are 363 and 861 for $\mathcal{C}_1$ and $\mathcal{C}_2$, respectively. The total length of input vectors, including $\bm{C}_j$ and $\bm{D}_j$, is $2(2\delta+1)=70$. For FBNet, both the forward and backward cells output a vector of length $70$, and the APP unit finally outputs a scalar value. For FBGRU, the size of the output for a bi-GRU module is 80, and the four fully-connected layers of the APP unit are designed as a 80$\times$40 layer followed by a 40$\times$20 layer, a 20$\times$10 layer, and a 10$\times$1 layer in the end. Moreover, {to accelerate the convergence of training, the weights of FBNet are initialized within specific ranges rather than random values according to the corresponding parameters in the FB algorithm. In particular,} $w_1$, $w_3$, $w_5$, $w_7$, $w_9$, and $w_{10}$ are set as 0.2, $w_2$, $w_6$, $w_{11}$, and $w_{12}$ are set as 0.6, and $w_4$, $w_8$, and $w_{13}$ are set as $-$6.0. Similarly, the weights $w_c$ and $w_d$ for FBGRU are set as $-$6.0, and the weights involved in the bi-GRU modules for FBGRU are initialized randomly.

Then, we implement both the proposed FBNet and FBGRU in PyTorch framework. The batch size is set as 20 for FBNet and 200 for FBGRU. The {number of epochs} is set as 300 for both FBNet and FBGRU. The two proposed neural networks are trained using training samples to minimize the binary cross-entropy loss via the Adamax optimizer, and the initial learning rate is set as 0.005.

In addition, all training and testing data are generated by the following procedure: 1) A binary information sequences $\bm{U}$ of length 191 (or 540) is randomly generated; 2) We perform LDPC encoder and marker insertion, and then the binary sequence $\bm{Y}$ of length 363 (or 861) is obtained; 3) The encoded sequence $\bm{Y}$ passes through the simulated channel defined by some given parameters, resulting in a received sequence $\bm{R}$. For the training of FBNet/FBGRU, we first generate sequences $\bm{Y}$ and $\bm{R}$ through the above process. Then the sequence pair $(\bm{R},\bm{Y})$ is formed as training samples, where $\bm{R}$ is the training data and $\bm{Y}$ is the label. For a set of different channel conditions, an equal number of pairs $(\bm{R},\bm{Y})$ are generated for each channel condition, and the pairs of different channel conditions are mixed to form the final training data set. The testing data set for a given channel condition is obtained by collecting 100,000 received sequences through the aforementioned process. We evaluate the bit error rate (BER) performances of various decoding methods on the testing data sets.

\subsection{Performance Comparisons for random insertions and deletions}
In this subsection, we conduct four experiments to illustrate error performances of the proposed FBNet/FBGRU and compare them with the original FB algorithm. In order to study the robustness of FBNet and FBGRU to CSI uncertainty, we consider a noisy estimate of CSI at receiver, i.e., the insertion and deletion probabilities $P_i$ and $P_d$ are corrupted by i.i.d. zero-mean Gaussian noise with variance $\delta_e^2$. In particular, we assume $P_i=P_d$ and $\delta_e=0.4P_d$ for convenience. The following two cases are considered in our simulations: 1) \textit{Perfect CSI}, in which the original FB algorithm has accurate knowledge of CSI, while FBNet and FBGRU are trained using labeled data generated with the same CSI used for generating the test data; and 2) \textit{CSI uncertainty}, in which the original FB algorithm is performed with the noisy version of CSI, and the labeled data used for training FBNet and FBGRU are generated with the noisy version of CSI. {It should be noted that the above simulation scenarios are similar to that for ViterbiNet \cite{Shlezinger20} and BCJRNet \cite{Shlezinger20-2,Tsai20}.}

\textit{\textbf{Experiment 1:}} In this experiment, we compare BER performances of the FB algorithm, FBNet, and FBGRU over ID-AWGN channels with CSI uncertainty. We consider $P_i=P_d=\{0.004,0.008,0.012,0.016,0.02\}$ for $\mathcal{C}_1$ and $P_i=P_d=\{0.002,0.004,0.006,0.008,0.01\}$ for $\mathcal{C}_2$. The signal-to-noise ratio (SNR) is set as 7.0dB for both $\mathcal{C}_1$ and $\mathcal{C}_2$ among all insertion/deletion probabilities.
For each channel condition, e.g., $P_i=P_d=0.004$ and SNR=7.0dB for $\mathcal{C}_1$, a testing data set of 100,000 sequences is generated with perfect CSI using the procedure described at the end of Section V-A.

We first consider the case of perfect CSI. According to the five different insertion/deletion probabilities, we generate two training data sets of size 200 and 20,000 with perfect CSI for FBNet and FBGRU, respectively. The FBNet is trained on the data set with 200 training samples, and then the training weights are used to simulate the BER performances over the five different testing data sets. Similarly, we train the FBGRU on the training data set with 20,000 samples, and the BER performances are evaluated over the same five testing data sets by using the training weights.

For the case of CSI uncertainty, the BER performances of FB algorithm are computed by using the noisy version of CSI over different testing data sets. For FBNet and FBGRU, the training data sets are generated similarly to the case of perfect CSI, except that the noisy version of CSI is used for ID-AWGN channel. The BER performances of FBNet and FBGRU are simulated through a training and testing procedure that is similar to the case of perfect CSI.

\begin{figure}[tb]\label{Fig12}
	\centering
	\includegraphics[height=2.5in]{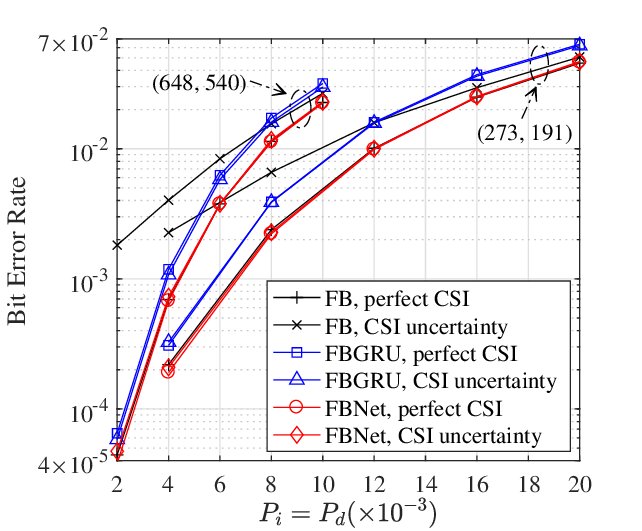}
	\caption{BER performance comparisons of various detection methods for $\mathcal{C}_1$ (273,191) and $\mathcal{C}_2$ (648,540) over ID-AWGN channels.}%
\end{figure}

Fig. 12 shows the BER performance comparisons of the original FB algorithm, FBNet, and FBGRU under the cases of perfect CSI and CSI uncertainty for both $\mathcal{C}_1$ and $\mathcal{C}_2$. It can be seen that the performances of the original FB algorithm with CSI uncertainty are deteriorated significantly when compared with that of the FB algorithm with perfect CSI. The BER performances of FBNet under the two cases of perfect CSI and CSI uncertainty can approach that of the FB algorithm with perfect CSI.
This validates the robustness of FBNet when working with CSI uncertainty.
Furthermore, it can be seen that FBGRU is robust to CSI uncertainty. However, there exist obvious performance gaps between FBGRU and the original FB algorithm with perfect CSI.

\begin{figure}[tb]\label{Fig13}
	\centering
	\includegraphics[height=2.5in]{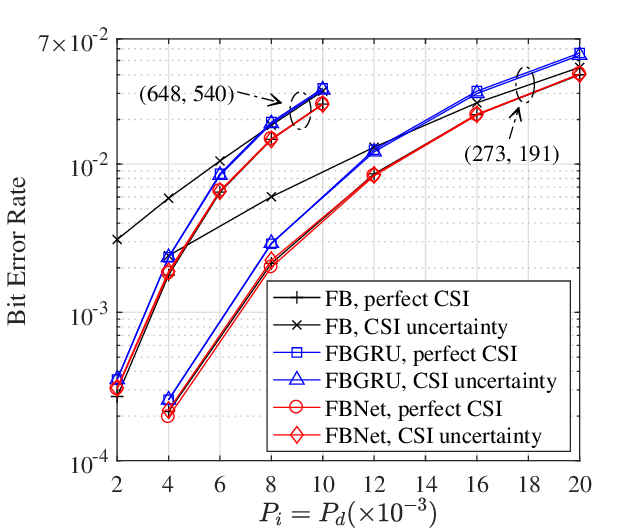}
	\caption{BER performance comparisons of various detection methods for $\mathcal{C}_1$ (273,191) and $\mathcal{C}_2$ (648,540) over IDS channels.}%
\end{figure}

\begin{figure}[tb]\label{Fig14}
	\centering
	\includegraphics[height=2.5in]{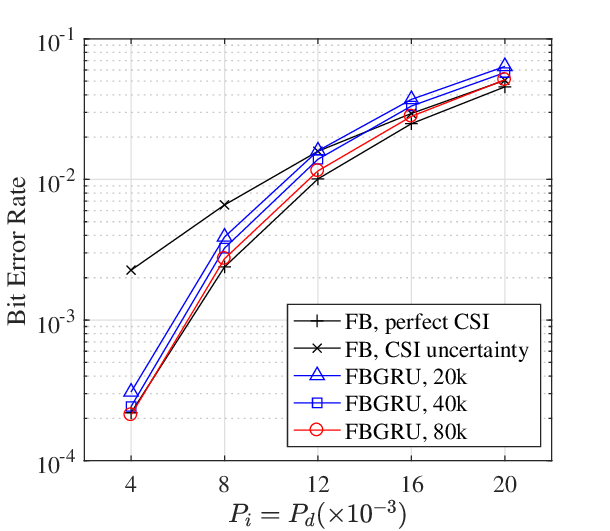}
	\caption{BER performance comparisons of FBGRU with different sizes of training data sets for $\mathcal{C}_1$ (273,191) over ID-AWGN channels.}%
\end{figure}

\textit{\textbf{Experiment 2:}} In this experiment, we compare BER performances of the proposed FBNet/FBGRU over IDS channels with CSI uncertainty. In the simulations, we set $P_s=0.004$ for both $\mathcal{C}_1$ and $\mathcal{C}_2$, and the values of $P_i$ and $P_d$ are the same as that in \textit{Experiment 1}. The training and testing data sets are also generated similarly to that in \textit{Experiment 1}. Then the FBNet and FBGRU are trained on the data sets of size 200 and 20,000, respectively. The BER performances of FBNet and FBGRU are simulated over the testing data sets. Fig. 13 shows the BER performances of various detection methods for $\mathcal{C}_1$ and $\mathcal{C}_2$ over IDS channels. Similar performance results can be observed as that for ID-AWGN channels. This further validates the robustness of FBNet and FBGRU for IDS channels.

\begin{figure}[tb]\label{Fig15}
\centering
\includegraphics[height=2.5in]{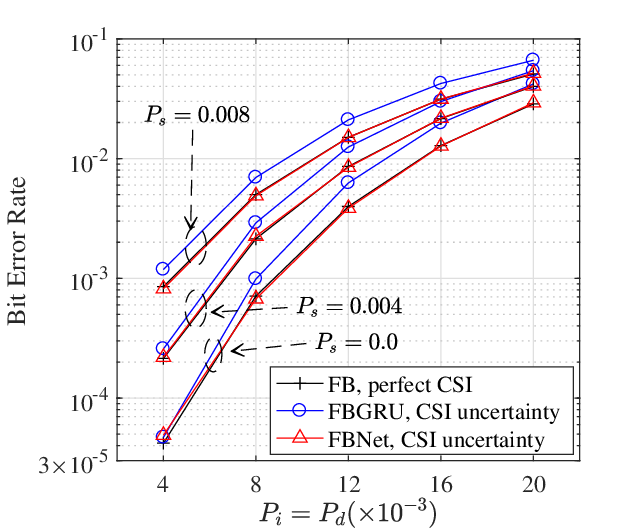}
\caption{BER performance comparisons of various detection methods for $\mathcal{C}_1$ (273,191) over IDS channels with different $P_s$.}%
\end{figure}
\begin{figure}[tb]\label{Fig16}
\centering
\includegraphics[height=2.5in]{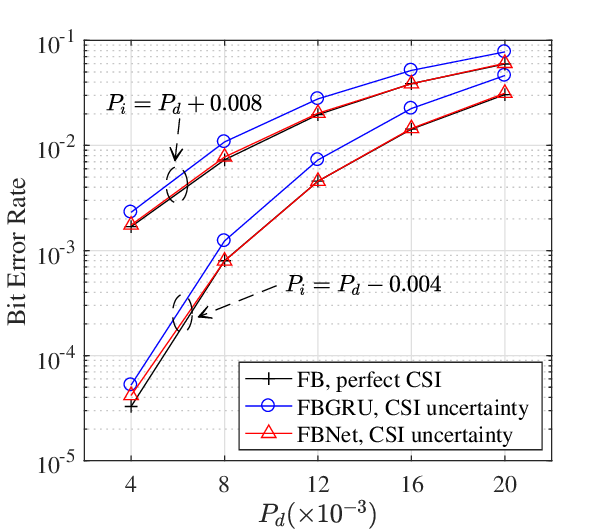}
\caption{BER performance comparisons of various detection methods for $\mathcal{C}_1$ (273,191) over IDS channels with $P_i \neq P_d$.}%
\end{figure}

\textit{\textbf{Experiment 3:}} For both ID-AWGN and IDS channels, there exist obvious performance gaps between FBGRU and the FB algorithm with perfect CSI in Experiments 1 and 2. In this experiment, {we want to figure out whether these performance gaps are caused by FBGRU itself or by insufficient training data.} For $\mathcal{C}_1$ over ID-AWGN channel with perfect CSI, we generate two additional training data sets of sizes 40,000 and 80,000 by five different insertion and deletion probabilities. Fig. 14 shows the BER performance comparisons of FBGRU with various training data sizes. {It can be seen that the performance gap between FBGRU and the FB algorithm with perfect CSI is gradually disappeared as the training data size increases.} Similar phenomena can be observed for $\mathcal{C}_2$ and IDS channels.

\textit{\textbf{Experiment 4:}} In this experiment, we investigate the robustness of the proposed FBNet and FBGRU over IDS channels with different $P_s$ and IDS channels with $P_i \neq P_d$. First, for $\mathcal{C}_1$ over IDS channel, we consider $P_s=\{0.0,0.004,0.008\}$ and $P_i = P_d=\{0.004,0.008,0.012,0.016,0.02\}$. For $P_s=0.0$, we generate the training and testing data sets following a similar approach as that in \textit{Experiment 1}. After training FBNet and FBGRU on training data sets with sizes of 200 and 20,000, respectively, the BER performances are evaluated through the five testing data sets. We repeat the above procedure for other values of $P_s$. Fig. 15 shows the BER performance comparisons of various detection methods for $\mathcal{C}_1$ over IDS channels with different $P_s$.

Second, for IDS channels with $P_i \neq P_d$, we set $P_d=\{0.004,0.008,0.012,0.016,0.02\}$ and consider two cases of $P_i=P_d-0.004$ and $P_i=P_d+0.008$, with $P_s=0.004$ kept unchanged. The data sets are generated similarly to other experiments and the BER performances are evaluated. Fig. 16 shows the BER performance comparisons of various detection methods over IDS channels with $P_i \neq P_d$. It can be observed from Fig. 15 and Fig. 16 that the proposed FBNet and FBGRU {with CSI uncertainty} also exhibit good performances on IDS channels with different $P_s$ or $P_i \neq P_d$. Similar phenomena can be observed over ID-AWGN channels.

\subsection{Performance Comparisons for weakly burst insertions and deletions}
In {the previous subsection}, the performances of FBNet and FBGRU are studied for synchronization channels with random insertions and deletions. However, the insertion and deletion events may not be independent in practice. For convenience, the channel may be modeled as random insertions and deletions at the receiver end. In this case, the performances of the FB algorithm will be degraded. Motivated by the described problem, we investigate the performances of FBNet and FBGRU with unknown synchronization channel models in this subsection. We first introduce a weakly burst insertion and deletion channel model. For the weakly burst IDS (WB-IDS) channel, the insertion, deletion, and substitution {events occur} with probabilities $P_{bi}$, $P_{bd}$, and $P_s$, respectively. When an insertion event occurs, $\kappa$ ($\kappa=2,3, {\rm or}\ 4$) consecutive random bits are inserted with probability $1/3$. When a deletion event occurs, $\kappa$ ($\kappa=2,3, {\rm or}\ 4$) consecutive bits are deleted with probability $1/3$. The weakly burst ID-AWGN (WB-ID-AWGN) channel can be described similarly. It should be noted that though the weakly burst synchronization channels may not be realistic, we mainly use them to evaluate the performances of FBNet/FBGRU with unknown channel models.

At the receiver, the detection algorithms, such as the FB algorithm, FBNet, and FBGRU, do not know the fact that the channel is weakly burst. The FB algorithm assumes insertions and deletions {occur} randomly and independent with each other. Moreover, considering that the mean of $\kappa$ is 3, the insertion and deletion probabilities used for the FB algorithm are set as $P_i=3P_{bi}$ and $P_d=3P_{bd}$, respectively. For FBNet and FBGRU, the training and testing data sets are generated according to the WB-IDS and WB-ID-AWGN channels.

\begin{figure}[tb]\label{Fig17}
	\centering
	\includegraphics[height=2.5in]{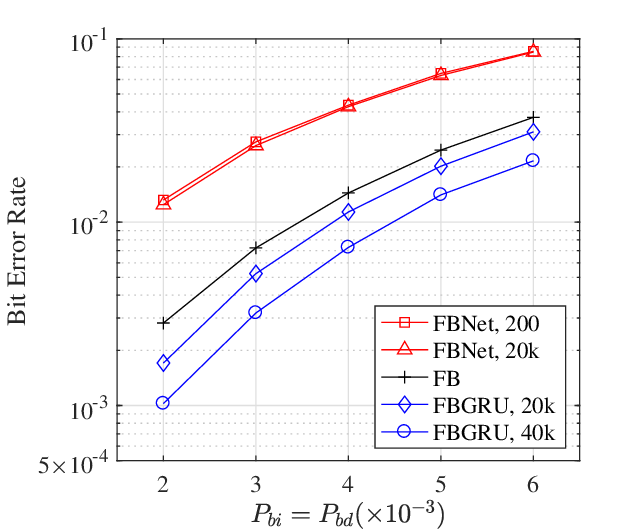}
	\caption{BER performance comparisons of various detection methods for $\mathcal{C}_1$ (273,191) over WB-ID-AWGN channels.}%
\end{figure}

\begin{figure}[tb]\label{Fig18}
	\centering
	\includegraphics[height=2.5in]{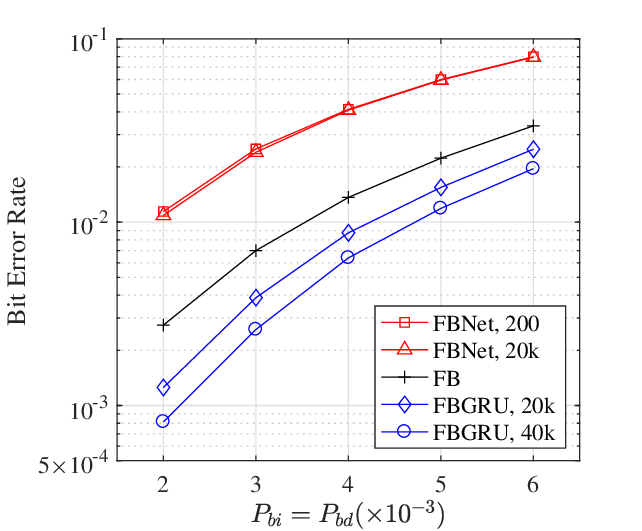}
	\caption{BER performance comparisons of various detection methods for $\mathcal{C}_1$ (273,191) over WB-IDS channels.}%
\end{figure}

\textit{\textbf{Experiment 5:}} In this experiment, the BER performances of FBNet and FBGRU are compared with the FB algorithm for $\mathcal{C}_1$ over WB-ID-AWGN channels. The SNR value is set as 7.0dB and $P_{bi}=P_{bd}=\{0.002,0.003,0.004,0.005,0.006\}$. For each $P_{bi}$, we generate 100,000 data sequences as testing data sets.
We generate three training data sets with sizes of 200, 20,000, and 40,000 according to the five different channel conditions. We train FBNet on training data sets of size 200 and 20,000, and evaluate the BER performances on the five testing data sets using the training weights.
After training FBGRU on data set of size 20,000, the BER performances are simulated on the five testing data sets. We also evaluate BER performances on the training data set of size 40,000 for FBGRU.

Fig. 17 shows the BER performance comparisons of FBNet, FBGRU, and the original FB algorithm for $\mathcal{C}_1$ over WB-ID-AWGN channels. It should be noted that the WB-ID-AWGN channel is only used to generate training and testing data sets and the receiver does not know the accurate channel model. It can be seen that the proposed FBGRU outperforms the original FB algorithm for the WB-ID-AWGN channel. This is because the FB algorithm, specially designed for random synchronization channels, is not optimal for weakly burst channels. The proposed FBGRU is a data-driven neural network and it has the capability of learning features from weakly burst channels. Moreover, as the size of the training set increases, the performances of FBGRU are significantly better than that of the FB algorithm. The BER performances of FBNet are significantly worse than that of the FB algorithm, and increasing the size of the training data sets does not lead to performance improvement. {The reason behind this is that FBNet is designed for random insertion/deletion channels and it neglects some events of consecutive insertions and deletions (see Eq. (\ref{Eq9})). It is possible to extend FBNet by incorporating additional computational cells to make FBNet perform better over WB-ID-AWGN/WB-IDS channels, and further studies are needed in this aspect. }

\textit{\textbf{Experiment 6:}} We also compare BER performances of the FB algorithm, FBNet, and FBGRU for $\mathcal{C}_1$ over WB-IDS channels, as shown in Fig. 18. The substitution probability $P_s$ is set as $0.004$. The training and testing data sets are generated similarly as that in \textit{Experiment 5}. Similar phenomena can be observed as that for the WB-ID-AWGN channel. This further validates the robustness of FBGRU over WB-IDS channels.

At last, we summarize the unique features of FBNet and FBGRU in Table I. For random insertions and deletions, FBNet is a good choice when accurate CSI is difficult to know since its fast training speed and small number of weights. However, FBNet is not suitable for unknown synchronization channel models such that the insertion and deletion events are not independent. In this case, FBGRU is a good choice since it is a data-driven method.

\begin{table}[!t]\label{table1}
	\centering
	\caption{A summary of the unique features of FBNet and FBGRU.}
	{
		\renewcommand{\arraystretch}{1.2}
		\begin{tabular}{|m{2.3cm}<{\centering}|m{2.4cm}<{\centering}|m{2.4cm}<{\centering}|}\hline
			& FBNet & FBGRU \\\hline
			Design basis & FB algorithm & bi-GRU \\\hline
			Driven mode & model-driven & data-driven \\\hline
			Number of weights & 13 & $\approx 10^{5}$ \\\hline
			Training data set & little (e.g., 200) & large (e.g., 20,000) \\\hline
			Applicability & random insertions and deletions & random insertions and deletions, unknown synchronization channel models\\\hline
		\end{tabular}
	}
\end{table}

\section{Conclusion}

In this paper, two deep learning-based detection methods FBNet and FBGRU have been proposed for marker code over synchronization channels to solve the performance degradation problem due to CSI uncertainty or unknown channel models. {FBNet is a model-driven method designed based on the original FB algorithm and it is robust to CSI uncertainty. FBGRU is a purely data-driven method designed based on bi-GRU. It performs better than FB algorithm and FBNet when the synchronization channel model is unknown at the receiver.}

{We end this paper by discussing several perspectives for future study. }

{1) Joint detection and decoding: In \cite{Tsai20}, the authors proposed a BCJRNet receiver for joint detection and decoding of Turbo codes over ISI channels. This is feasible because both detection and decoding use BCJR algorithms and these two BCJR modules can be merged into a single BCJR receiver. For marker codes, integrating FB detection and LDPC decoding into a single algorithm needs additional efforts. Therefore, designing a joint detection and decoding algorithm for marker codes and its robust version based on deep learning deserve further study.}

{2) Transformer-based receiver: Transformer has achieved great success in many artificial intelligence fields. It was originally proposed as a sequence-to-sequence model for machine translation. Since the detection of marker codes can be seen as a sequence prediction problem, transformer may be used as a substitute for bi-GRU. In addition, transformer can also be used to design LDPC decoder with improved performances \cite{Choukroun22}. Therefore, it is an interesting research problem to design a single transformer for the detection and decoding of marker codes.}




\end{document}